\documentclass[preprint1]{aastex6}

\usepackage{soul}\usepackage{graphicx}   
\usepackage{dcolumn}    
\usepackage{bm}         
\usepackage{verbatim}   
\usepackage{booktabs}
\usepackage{amsmath,amssymb}
\usepackage{color}
\usepackage{multirow}
\usepackage{inputenc}
\usepackage{natbib}
\bibpunct{(}{)}{;}{a}{}{,}
\usepackage{floatrow}
\usepackage[caption=false]{subfig}
\usepackage{savesym}
\savesymbol{tablenum}
\usepackage{siunitx}
\restoresymbol{SIX}{tablenum}
\usepackage{enumitem}   

\usepackage[normalem]{ulem}

\newcommand{\aver}[1]{\ensuremath{\left\langle#1\right\rangle}}

\newcommand{\beq}{\begin{equation}}
\newcommand{\eeq}{\end{equation}}
\newcommand{\bea}{\begin{eqnarray}}
\newcommand{\eea}{\end{eqnarray}}
\newcommand{\p}{\partial}
\newcommand{\mb}{\mathbf}

\begin{document}
\title{Particle Acceleration and Fractional Transport in  Turbulent Reconnection}
\author{ Heinz Isliker, Theophilos Pisokas, Loukas Vlahos}
\affiliation{Department of Physics, \; Aristotle University of Thessaloniki\\
GR-52124 Thessaloniki, Greece}
\author{Anastasios Anastasiadis}
\affiliation{Institute for Astronomy, Astrophysics, Space Applications and Remote Sensing,\\
National Observatory of Athens\\
GR-15236 Penteli, Greece}

\date{\today}
\begin{abstract}
	We consider a large scale environment of turbulent reconnection that is fragmented into a number of randomly distributed Unstable Current Sheets (UCS), and we statistically analyze the acceleration of particles 
	within this environment.
	We address two important cases of acceleration mechanisms when the particles interact with the UCS: (a) electric field acceleration, and (b) acceleration through reflection at contracting islands. Electrons and ions are accelerated very efficiently, attaining an  energy distribution of power-law shape with an index $1-2$, depending on the acceleration mechanism. 
	The transport coefficients in energy space 
	are estimated from the test-particle simulation data, 
	and we show that the classical Fokker-Planck (FP) equation fails to reproduce the simulation results
	when the transport coefficients are inserted into it and it is solved numerically. The cause for this failure is that the particles perform Levy flights in energy space, the distributions of energy increments exhibit power-law tails. We then use the fractional transport equation (FTE) derived by Isliker et al., 2017, whose parameters and the order of the fractional derivatives are inferred from the simulation data, and, solving the FTE numerically, we show that the FTE successfully reproduces the kinetic energy distribution of the test-particles. 
	We discuss in detail the analysis of the simulation data and the criteria 
	that allow judging the appropriateness of either an FTE or a classical FP equation as a transport model. 
\end{abstract}

\maketitle
\section{Introduction}
A few years ago it was believed that the prominent acceleration mechanisms for space and astrophysical plasmas were: (1) The large scale DC electric field associated mostly with a reconnecting current sheet, (2) a spectrum of weak amplitude MHD waves and (3) the diffusive shock acceleration \cite[see reviews by][]{Miller97, Melrose94}. Which mechanism will dominate in the acceleration of the charged particles in space, astrophysical or laboratory plasmas was assumed to be related with the global energy release process(es) and the magnetic topology. The diffusion of particles in the six dimensional position and momentum space is a complex problem in astrophysical and laboratory plasmas. In order to avoid this complexity, most studies concentrated their analysis on the energetics of the interaction of particles with isolated nonlinear MHD structures, e.g.\  reconnecting current sheets, a spectrum of small amplitude MHD waves, and shocks.

The reconnecting current sheet was analyzed for simplicity mostly as an isolated 2D  or 3D structure \citep{Priest14}. The processes related to the reconnection were assumed to be laminar and the field lines smooth and well behaved. The conditions on the inflow and outflow sides of the reconnection zone have been taken, again for simplicity, to be quiescent. The structure was assumed to be ``long'' lived and ``stable''. The electric field associated with this highly idealised topology was approximated as $\mb{E} \approx -\mb{V}_A \times \mb{B}/c+\eta \mb{j}$, were $\mb{V}_A$ will be close to the Alfv\'en speed, $\mb{B}$ is the ambient magnetic field, $c$ the speed of light, $\eta$ is the resistivity, and $\mb{j}$ the current density inside the reconnecting volume. The presence of broad band fluctuations of the magnetic and velocity fields, or the simultaneous presence of multiple reconnecting current sheets, or the generalization to a 3D magnetic topology, will change the above simplified picture and lead to a completely new scenario for magnetic reconnection  where current fragmentation will lead to strong turbulence $(\delta B/B \approx 1)$. In a series of recent reviews the  road  from the reconnecting current sheet(s) to the random formation of an ensemble of unstable current sheets (UCSs) inside a strongly turbulent plasma was outlined \cite[e.g.][and references therein]{Matthaeus11, Hoshino12, Cargill12, Lazarian12, Karimabadi13, Karibabadi2013c}. This new environment of multiple UCSs was named {\bf turbulent reconnection}. How electrons and ions will be accelerated by large scale turbulently reconnecting plasmas in numerous space, astrophysical or laboratory settings remains an open problem, and we will return to this issue in the next section.

The stochastic turbulent acceleration of charged particles by a spectrum of weak amplitude MHD waves was analyzed initially by \cite{Kulsrud71} and \cite{Krymskii77}, who followed closely the original idea of \cite{Fermi49}. The proposed mechanism was based on the assumption that the magnetic filed topology is simple and the excited MHD modes will be in resonance with the ions and/or electrons. The cause of the excitation of the MHD waves was only {\bf schematically} connected to the energy release process(es) and the fact that the waves were saturated at small amplitudes ($\delta B/B <<1$) was never justified. The main assumptions on which stochastic wave-particle interaction was based are (a) the level of the excited waves should remain low for the quasilinear assumption to be valid and (b) the Fokker Planck equation will also be valid \citep{Achterberg81, Miller90, Petrosian12}. 
The break of these assumptions leads to a very different scenario for turbulence and particle acceleration \citep{Biskamp89, Galsgaard96, Dmitruk04, Arzner04, Turkmani05, Arzner06}. Therefore a second road from {\bf strong turbulence to turbulent reconnection} was established in several 2D or 3D simulations the last thirty years (see the review of \cite{Matthaeus11} and the cited literature). 

The analysis of the turbulent shock followed the last fifty years the same simplifications outlined above for the reconnecting current sheet and the weak MHD waves. The analysis of the shock and the acceleration of particles was always done in idealized topologies. The main scenario was based on a 2D smooth magnetic discontinuity (shock front) which moves with velocity $V$ larger than the Alfv\'en speed inside a spectrum of MHD waves, which are convected with the flows upstream and downstream of the magnetic discontinuity. The charged particles are trapped and bounce from upstream to downstream of the shock front several times till they will escape from the vicinity of the shock. In the idealized case two mechanisms can accelerate particles at the shock, the shock drift acceleration, which dominates in quasi-perpendicular shocks, and the diffusive shock acceleration, which dominates in quasi-parallel shocks \citep{Decker88, Drury83, Kirk01, Schurel12, Lee2012}. The presented picture changes dramatically when the turbulence upstream and downstream in a 3D magnetic structure reaches very large amplitudes ($\delta B/B \approx 1$) \citep[see][]{Caprioli14a, Caprioli14b, Caprioli14c} and drives upstream and downstream of the shock front turbulently reconnecting plasmas \citep{Lazarian09, Drake10, Karimabadi2014, Matsumoto15, Zank15,Burgers16}. How the new system of a shock with turbulently reconnecting plasmas upstream and downstream will evolve and how the turbulent flows may eventually transform the turbulent shock to a new environment full of randomly moving small scale internal shock waves and UCSs \citep{Achterberg90b, Anastasiadis91, Anastasiadis94,   Schneider93} remain open problems. In other words, shock waves seem to follow the same pattern as the UCS(s) and  strong MHD turbulence, asymptotically they end up in a turbulent reconnection environment \cite[e.g.][and cited references]{Karimabadi2014}

In summary, all three prominent acceleration mechanisms listed above: (a) magnetic reconnection, (b) strong MHD turbulence $(\delta B/B \approx 1)$, and (c) shock waves, will evolve towards a new accelerator that has received considerable attention the last ten years and is called {\bf turbulent reconnection}.  

In this article we analyze the evolution of the energy distribution function of electrons and ions, their escape time, the role of collisions, and the transport properties of the particles inside 
a turbulent reconnection environment, where
an ensemble of {\bf localized} and highly effective accelerators is present. Our simulation box is large and we follow the system for long times that are close to the observational ones, contrary to the current Particle In Cell (PIC) simulations. The local accelerators that we consider may be UCSs, magnetic islands, or internal shocks, and they will systematically accelerate  particles as they cross the turbulently reconnecting volume. The interaction of the charged particles with the small scale UCS or internal shocks follows the scenario proposed by \cite{Fermi49, Fermi54}, with the magnetic clouds being replaced by the more general concept of active {\bf scatterers}. Particular aim of this study also is to explore the possible limitations of the FP approach in modeling transport, and to show the importance of fractional transport equations in the context of turbulent reconnection.

In Section \ref{s:particle}, we analyze the local dynamics of particles in the interaction with typical scatterers (i.e.\ UCSs). 
In Section \ref{s:class_frac}, we introduce a fractional approach for modeling transport, in comparison to the classical, Fokker-Planck approach
and the classical transport coefficients.
In Section \ref{s:numerical}, we construct a 3D lattice gas model, where a small number of grid sites are {\bf active (scatterers)}, and we study the energy evolution of an initial distribution of particles injected into this environment. In Section \ref{s:transport}, we examine the transport properties of the particles inside a collection of UCSs, using both, the fractional and the FP approach. In the final section we discuss the implications of our results for explosive phenomena in the solar corona and list our main conclusions.

\section{Particle acceleration by an ensemble of Unstable Current Sheets}\label{s:particle}
\cite{Matthaeus84} and \cite{Ambrosiano88} were the first to analyze numerically the test particle dynamics inside a 2D {\bf turbulently reconnecting} environment, using the electromagnetic fields 
form the simulations of \cite{Matthaeus80} and \cite{Matthaeus86}. These original studies analyze the role of turbulent reconnection in the acceleration of test particles. Several studies returned to this problem many years later to analyze the interaction of particles in 3D topologies, where the turbulently reconnecting environment was generated through large amplitude waves or multiple UCSs, using the MHD equations as the main tool for the understanding of the current fragmentation and the formation of turbulently reconnecting environments \citep{Dmitruk03, Dmitruk04, Arzner04, Turkmani05, Onofri06, Arzner06, Kowal11, Gordovskyy11}. The resistive MHD equations can handle the large scales in space, astrophysical, and laboratory systems and they can follow the formation of UCS, but they miss completely the kinetic evolution. Also, the statistical properties of the UCSs inside a 2D or 3D turbulently reconnecting environment has been analyzed by \citet[][]{Uritsky10, Servidio11, Zhdankin13}.

On different scales, the analysis of the UCS(s) through the use of Particle In Cell (PIC) codes focuses on the details of the dynamics of the particles. They follow the formation of plasmoids and their interaction with electrons and ions \citep{Onofri06, Drake06, Guo2014, Guo15, Dahlin15, Hoshino12b}. The PIC simulations, both in 2D or 3D, use periodic boundary conditions, the dimensions of the simulation box are extremely small (several meters for solar coronal parameters), and the time scale of their evolution is a few thousands of $\omega_e^{-1},$ where $\omega_e$ is the electron plasma frequency (for the solar corona this is several microseconds). Thus, the PIC simulations follow the kinetic aspects of turbulent reconnection in a very small periodic volume and for very short times. 

It is important to separate the particle dynamics inside evolving UCS(s) from the evolution in the global environment where particles travel between UCSs, which serve as scatterers, in analogy with the {\bf magnetic clouds} envisioned by \cite{Fermi49}. Particles travel a distance $\lambda_\textrm{sc}$ before they interact with an UCS, and the energy gain, as they ``cross'' the scatterer, is $dW$. The energy increment ($dW$) can be stochastic \cite[see][]{Pisokas16} when the particles interact with magnetic fluctuations, or systematic ($dW>0$) when they interact with UCSs: Based on current PIC simulations, the energy increments, when the particles escape from the locally evolving UCS, are always positive (see \cite{Guo15, Dahlin15, Matsumoto15}). 

\cite{Vlahos04, Vlahos16} and \cite{Pisokas16} attempted to combine the large scale dynamics of particles with their dynamics at the UCS, assuming a very simple interaction of the particles with the locally evolving UCS \cite[see also the recent reviews][]{Cargill12, Lazarian12, Karimabadi13, Karimabadi03b}. It was assumed that the spatial and temporal scales of the UCSs are so small compared with the system under study that they can be approximated with nodes inside the large scale simulation box, and that their interaction with the particles is instantaneous.

\subsection{Systematic acceleration by random electric fields}\label{ss:EFtheor}
The particle dynamics inside the UCS is complex, since internally the UCS are also fragmented and the particles that interact with the fragments of the UCS can lose and gain energy on the microscopic level of description. Yet, on the average and over the entire simulation domain, the particles gain energy systematically before exiting the UCS, see Fig.~6(c) of \cite{Guo15} or Fig.~3d in \cite{Matsumoto15} and the related discussion. 
We approximate the {\bf macroscopic} energy gain as
\begin{equation}\label{e:dW_ef}
    \Delta W = |q| E_\textrm{eff}\ \ell_\textrm{eff},
\end{equation}
where $E_\textrm{eff} \approx (V/c) \, \delta B$ is the measure of the effective electric field of the UCS, and $\delta B$ is the fluctuating magnetic field encountered by the particle, which is of stochastic nature, as related to the stochastic fluctuations induced by reconnection. $\ell_\textrm{eff}$ is the characteristic length of the interaction of the particle with the UCS and should be proportional to $E_\textrm{eff}$, since small $E_\textrm{eff}$ will be related to small scale UCS. In other words, the UCS is a complex structure which always accelerates the charged particle crossing it. The scenario of the method used here is: particles approach the scatterers with an initial kinetic energy $W_0$ and depart with an energy $W = W_0 + \Delta W$, where the energy gain $\Delta W$ {\bf on the macroscopic level}  is systematic and follows the statistical properties of the fluctuations $\delta B$.

\subsection{Systematic Fermi acceleration at contracting islands}\label{ss:SFtheor}

The systematic and macroscopic energy gain from the acceleration through reflection at contracting islands in an UCS can be written as a variant of the classical Fermi acceleration for relativistic particles
when only head on collisions are taken into account and the energy increase is always positive,
\begin{equation}\label{e:dW_sf}
    \Delta W = \Gamma \left [ \left (\frac{V_A}{c} \right )^2 + \left ( \frac{2V_A u_\parallel }{c^2} \right ) \right ] W
\end{equation}
\citep{Longair11}, where $u_\parallel$ is the component of the velocity of the particle parallel to the magnetic field and $\Gamma$ is the Lorenz factor. For nonrelativistic particles we have
\begin{equation} \label{e:dW_nrs}
    \Delta W = W \left (\frac{V_A}{c} \right ) \left ( \frac{u_\parallel }{c} \right ) = W \left (\frac{V_A}{c} \right ) \left ( \frac{u}{c} \right ) \cos \phi
\end{equation}
where $\phi$ is a random angle, $-\pi/2<\phi<\pi/2.$ \citep{delPino05, Drake06, Hoshino12b, Dahlin15, Guo15}.

The important difference between the systematic electric field ``scatterer'' and the systematic Fermi acceleration is the fact that $dW$ is not a function of $W$ in the first case, in contrast to the second case. 



\section{Classical and fractional approach to modeling transport}
\label{s:class_frac}

\subsection{Classical transport coefficients and the Fokker Planck equation}\label{ss:FPtheor}

	The energy convection coefficient, which reflects the systematic aspects of
	acceleration, is given as
	\begin{equation}\label{eq:FW}
	F(W,t) =\frac{\aver{W(t+\Delta t) - W(t)}_W}{\Delta t},
	\end{equation}
	and the energy diffusion coefficient, which describes the statistic nature of acceleration, is defined as
	\begin{equation}\label{eq:DWW}
	D(W,t) = \frac{\aver{\left(W(t+\Delta t) - W(t)\right)^2}_W}{2\Delta t},
	\end{equation}
	where $\aver{\ldots}_W$ denotes the conditional average that $W(t) = W$, which accounts for the functional dependence of the transport coefficients on the energy $W$ \cite[see e.g.~][]{Ragwitz2001}. 
	In the current literature these coefficients were often estimated analytically with the use of several simplifying assumptions \citep{Drake13, Guo2014, Zank15}, whereas we here determine them directly from the particle dynamics, as in \cite{Vlahos16}.
	For the numerical estimate of the coefficients, the energies of the particles at time $t$ are divided into logarithmically
	equi-spaced bins, and the coefficients are estimated for each bin separately (i.e.\ we apply binned statistics). $\Delta t$ is a small time-interval, just large enough so that most particles exhibit discernible changes of the their kinetic energy (the theoretical limit $\Delta t \to 0$ can of course practically not be applied).

	
	As as simplification, we here neglect spatial diffusion and consider that the particles diffuse only in energy space. The well established assumption in space and astrophysical plasmas is that the interaction of particles with the UCSs is such that the Fokker Planck (FP) equation will be valid,
	\begin{equation} \label{diff}
	\frac{\partial n}{\partial t} +
	\frac{\partial}{\partial W} \left[F n -\frac{\partial (D n) }{\partial W} \right] =
	-\frac{n}{t_{\rm esc}} + Q ,
	\end{equation}
	where $n$ is the energy distribution function, $t_{\rm esc}$ is the escape time from an acceleration volume with characteristic size $L$, $Q$ is the injection rate, and $F$ and $D$ are the transport coefficients defined above. The validity of the FP equation is based on the hypothesis that the particles execute a kind of Brownian motion in the energy space as they interact with the localized UCSs, with the random walk steps being small in the sense that they have finite mean and variance (see the remarks below). In section \ref{s:transport} we use data from the test particle approach 
	to asses the validity of the FP equation 
	in a turbulently reconnecting plasma.

	Using the pseudospectral method 
	with rational Chebyshev polynomials as base functions in energy space, we numerically integrate the FP equation 
	on the semi-infinite energy interval $[0,\infty)$, 
	with the time-stepping being implemented
	in the form of the implicit backward Euler method (see e.g.\ \cite{Boyd01}).


\subsection{Fractional transport equation (FTE)}
\label{ss:FTEtheor}
 
Fractional transport equations allow describing non-local and anomalous transport phenomena. In the derivation of the FTE we here follow the one given in \cite{Isliker17}.  
A variant of the Chapman-Kolmogorov equation, 
\beq
n(W,t) = \int dw\int_{0}^{t} d\tau\,n(W-w,t-\tau) \, q_{w }(w) 
\, q_\tau(\tau) 
+ n(W,0)\int_{t}^{\infty}q_\tau(\tau) d\tau ,
\label{chapkol}
\eeq
allows for
a general description of transport in energy space, 
see e.g.\ \cite{Klafter87,Klages08}. The conservation law in Eq.\ (\ref{chapkol}) 
can be interpreted as a Continuous Time Random Walk (CTRW) process
(a generalization of the classical Brownian motion). 
The random walk steps in energy $w$ are distributed according to the 
the probability density $q_w$, and the time 
intervals $\tau$ for these steps to be performed obey
the probability density function $q_\tau$ (the two probability densities are assumed to be independent in order to simplify the approach). When both distributions of increments, $q_w$ and $q_\tau$,  
have finite mean and variance (as e.g.\ in the case that they are Gaussian distributions), then, by a standard procedure through Taylor expansions, the FP equation (Eq.\ (\ref{diff})) can be derived from Eq.\ (\ref{chapkol}),
see e.g.\ \cite{Gardiner09}. In the following, we do not make the 
assumption that the increments are small and explicitly allow for infinite mean or variance.

In order to derive a meso-scopic equation, we make 
a Fourier Laplace transform ($ W\to k $, $t\to s$) of Eq.\ (\ref{chapkol}) and make use of the respective convolution theorems, which yields
the Montroll-Weiss equation (written in slightly nonstandard form),
\beq
\tilde{\hat n}(k,s) = \tilde{\hat n}(k,s) \, \hat q_{w }(k) 
\, \tilde q_\tau(s) + \hat n(k,0) \frac{1 - \tilde q_\tau(s)}{s} 
\label{chapkolFL}
\eeq
\citep{Montroll65,Klafter87}.

As distributions of increments we employ the stable Levy distributions, which are defined in Fourier ($k$) or Laplace ($s$) space through their so-called characteristic functions.
For the distribution of energy increments, we 
assume the symmetric stable Levy distributions
$\hat   q_w(k) = \exp(-a|k|^\alpha)$,
with $0<\alpha\leq 2$, which have a power-law tail in energy-space,
$q_w(w)\sim 1/w^{1+\alpha}$ for $\alpha<2$ and $w$ large, and 
for $\alpha=2$ 
they are Gaussian distributions \citep{Hughes95}.
For the waiting time distribution, we consider one sided stable Levy distributions (expressed in Laplace space),
$\tilde q_\tau = \exp(-bs^\beta)$
with $b>0$ and $0<\beta\leq 1$, which exhibit a power-law tail, 
$q_\tau \sim 1/\tau^{1+\beta}$ for $\beta<1$ and $\tau$ large,
and for $\beta = 1$ they are Dirac delta-functions, $q_\tau(\tau)=\delta(\tau-b)$ \citep{Hughes95}. 

The last step in the derivation of the FTE consists in applying the fluid-limit, i.e.\ we assume that   
$w,\,\tau$ are large, and thus in turn that $k,\,s$ are small, (e.g.\ \cite{Klages08}, and references therein), which allows approximating 
the distributions of increments by a Taylor expansion up to first order, 
$\hat   q_w \approx 1-a|k|^\alpha$ and
$\tilde q_\tau \approx 1 - bs^\beta$.
Inserting the approximations into Eq.\ (\ref{chapkolFL}) yields
\beq
\tilde{\hat n}(k,s) = \tilde{\hat n}(k,s) \, (1-a|k|^\alpha -bs^\beta) + 
\hat n(k,0) bs^{\beta-1} ,
\label{chapkolFL_fl2}
\eeq
or, by rearranging,
\beq
bs^\beta \tilde{\hat n}(k,s) - bs^{\beta-1} \hat n(k,0)  = - a|k|^\alpha \tilde{\hat n}(k,s)   .
\label{chapkolFL_fl}
\eeq
The latter equation actually contains fractional derivatives
and can be written as a fractional transport 
equation
\beq
b D_t^\beta n = a D_{|W|}^\alpha n ,
\label{fract}
\eeq
where 
$D_t^\beta$ is the Caputo fractional derivative of order $\beta$,
defined in Laplace space as
\beq
{\cal L}\left(D_t^\beta n\right) = s^\beta \tilde n(W,s) -s^{\beta -1} n(W,0)
\eeq
and $D_{|W|}^\alpha$ is the symmetric Riesz fractional derivative of order $\alpha$,
defined in Fourier space as
\beq
{\cal F}\left(D_{|W|}^\alpha n\right) = -|k|^\alpha \hat n(k,t) ,
\eeq
see e.g.\ \cite{Klages08,Bian08}.
Note that for $\beta=1$ and $\alpha=2$ or $1$, Eq.\ (\ref{fract}) includes the cases of a pure diffusion or convection equation, respectively, 
as corresponding to the classical Fokker-Planck equation. 

The derivation of the FTE makes it clear that
the indices of the power-law tails, if any, of the distributions of increments $q_w(w)$ and $q_\tau(\tau)$ determine
the order of the fractional derivatives. On the other hand,  
if there are no power-law tails and both the mean and variance of the distributions of increments are finite, then the respective derivatives are classical and of integer order. Finally, if both the time and energy derivatives are of integer order, then  
the classical FP equation is appropriate. 

The parameters of the FTE are $\alpha$, $a$ and  $\beta$, $b$,
which need to be estimated from the test-particle data.
As pointed out above, $\alpha$ 
is given by the index $z$ of the power-law tail of $p_w(w)$
as $\alpha = -z-1$. The characteristic function approach
\citep{Borak05,Koutrouvelis80} is a second method to determine $\alpha$ and also $a$. Based on the sample of increments $\{w_j\}$ from the test-particle simulations, the estimator $\tilde{q}_w$ of the characteristic function $\hat{q}_w$ is defined as
\beq
\tilde{q}_w(k) = \langle e^{ik w_j}\rangle_j
\label{e:char_fun}
\eeq
for a suitable set of $k$-values. If the $w_j$ are distributed according to a stable Levy distribution,
then $\hat   q_w(k) = \exp(-a|k|^\alpha)$ should hold, and $\alpha$ will equal the slope of a linear fit to 
$\ln (-\ln |\tilde{q}_w|^2)$ as a function of $\ln k$,
and the intercept with the $y$-axis will yield $\ln(2 a)$. 

Before turning to the temporal part of the FTE and its parameters $\beta$ 
and $b$, we need to specify how we define the energy increments $w_j^n$ (with $j$ the particle index and $n$ the index that labels temporally subsequent increments). In the lattice model introduced below, the energizations are discrete events that take place in the instantaneous interactions of the particles with the scatterers. So one choice could be to let $w_j^n$ denote these instantaneous changes of particle energy. This choice though has practical difficulties in its application e.g.\ to test-particle simulations, as e.g.\ in \cite{Isliker17}, where the energizations may or may not be localized, and where very likely there would be some arbitrariness in the definition of the waiting times. A second choice, which, as in \cite{Isliker17}, we adopt here, is to monitor the particles at fixed time intervals of duration $\Delta t$, and to consider the increments to be defined as the energy change over the time-interval $\Delta t$, so that the increments $w_j^n$ become      
equal to the energy differences $[W(t+\Delta t) - W(t)]_j$ in the definitions of $F$ and $D$, Eqs.\ (\ref{eq:FW}) and(\ref{eq:DWW}),  
\beq
w_j^n \equiv w_j(t) := [W(t+\Delta t) - W(t)]_j .
\label{delta_w}
\eeq

The definition of increments $w_j^n$ used here then implies that the waiting times are constant ($\Delta t$), 
and thus the waiting time distribution is of the form $p_\tau(\tau)=\delta(t-\Delta t)$, from which it follows, as explained above, that $\beta=1$ and 
$b=\Delta t$. 
Thus, in the following we consider the fractional transport equation 
to be of the form
\beq
\p_t n = (a/\Delta t) D_{|W|}^\alpha n - n/t_{esc} ,
\label{fracth}
\eeq
with an ordinary, first order derivative in time-direction and a fractional derivative in energy direction,
and where we also have added the escape term $- n/t_{esc}$.

We solve the fractional transport equation numerically by using the Gr\"unwald-Letnikov definition
of fractional derivatives (e.g.\ \cite{Kilbas06}), implemented
in the matrix form 
of \cite{Podlubny09}, and in particular we use 
the derivative scheme of \cite{Podlubny13} for non equi-distant grid-points, which allows using the same grid-points in $[0,\infty)$ as for the solution of the FP equation above. In time direction, the implicit backward Euler scheme is used. 
Since the FTE has been derived here as a tool for modeling long tails at the high energy side of the energy distribution, which is the main interest of this study,
we apply the fractional derivative only above energies of $10\,$eV in the numerical applications (note, e.g., that the fluid limit has been applied in the derivation of the FTE).



\section{Particle dynamics in turbulently reconnecting plasma: numerical study}\label{s:numerical}

\subsection{Initial set-up}\label{ss:setup}

We use a lattice gas model for the simulation of turbulent reconnection,
in the form of a 3D grid $(N \times N \times N)$ with grid size $\ell=L/(N-1)$ and linear extent $L$. A randomly chosen, small 
fraction $R = N_{\rm sc}/N^3$ (5-\SI{20}{\percent}) of grid-sites
is marked as \emph{active scatterers}, the rest of the grid-sites is inactive. 
The mean free path of the particles between scatterers can be determined as $\lambda_{\rm sc}=\ell/R,$ 
and the density of the scatterers follows as $n_{\rm sc} = R \times N^3/L^3$. 

At time $t = 0$, a large number of particles (electrons or ions) 
is distributed in the simulation box over randomly chosen grid-sites.  
The particles initially follow a velocity distribution $n(W, t=0)$ of  Maxwellian shape with temperature $T$, 
and they move into random directions on the grid
(the particles are bound to follow the grid-lines).

	\begin{figure}[ht] 
		\sidesubfloat[]{\includegraphics[width=0.40\columnwidth]{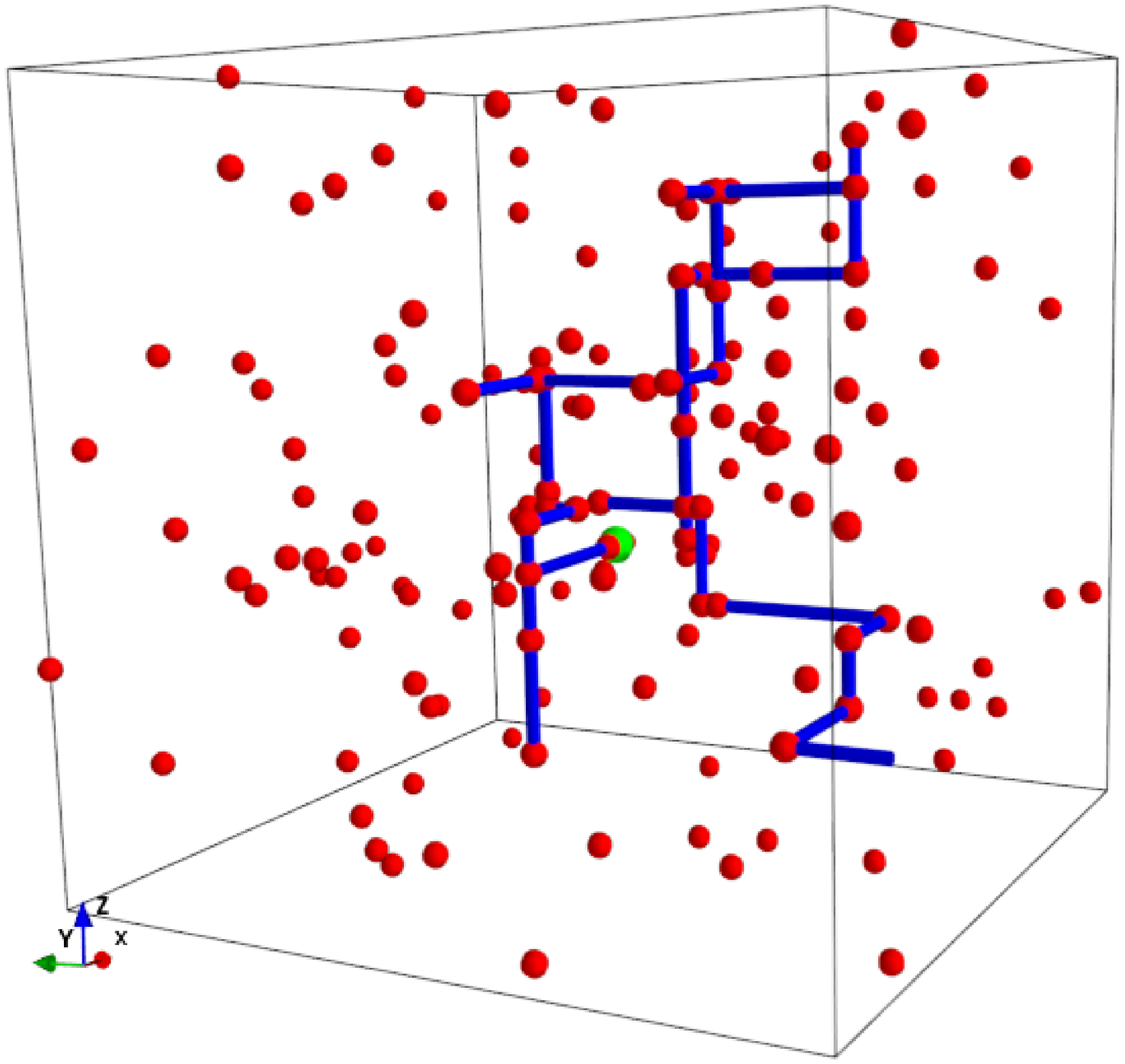}%
			\label{f2a}}\hfill%
		\sidesubfloat[]{\includegraphics[width=0.45\columnwidth]{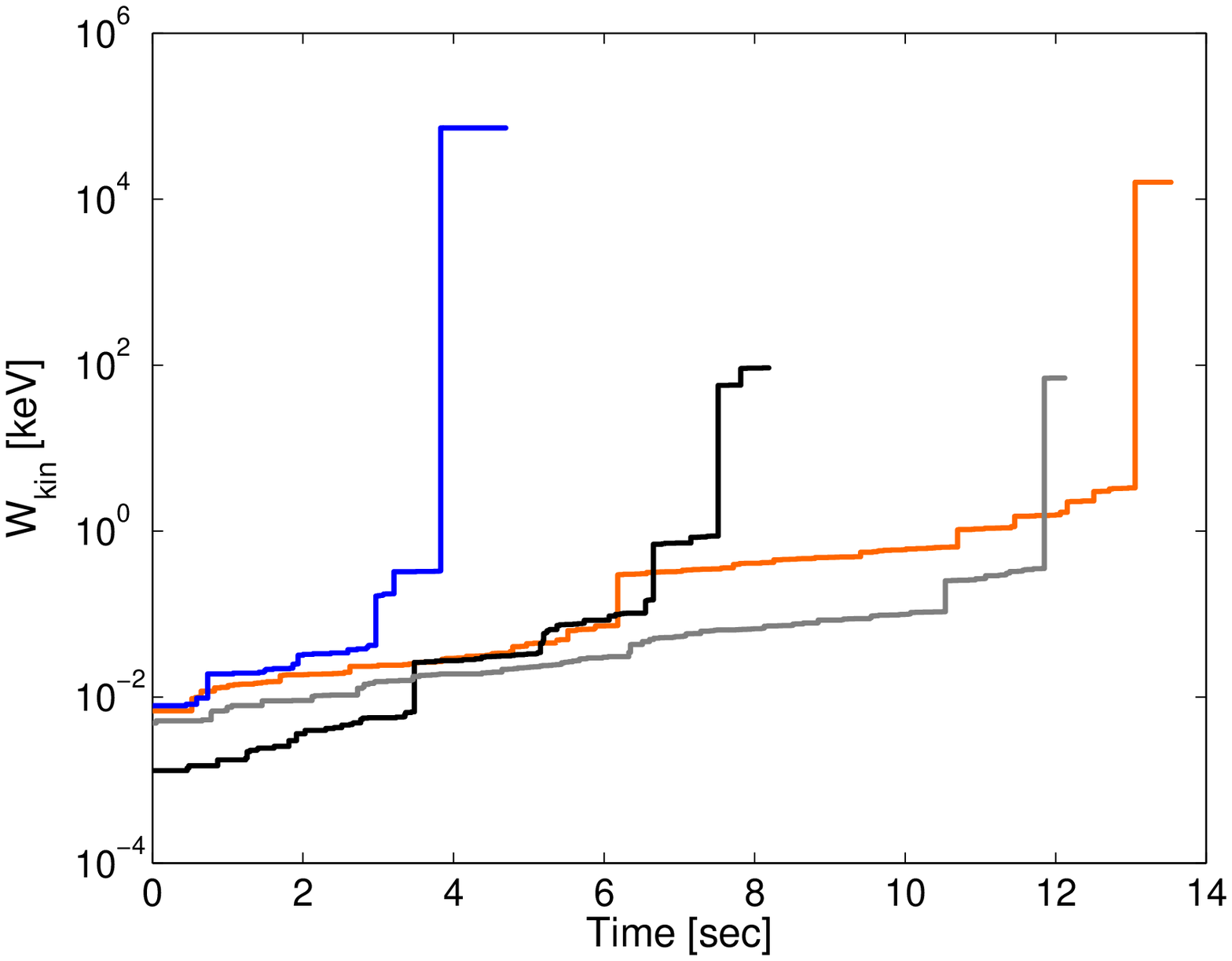}%
			\label{f2b}}%
		\caption{ (a) A typical trajectory in the 3D simulation box (blue). The  red spheres mark the randomly distributed  UCSs (active grid-sites). (b)  The energy as a function of time for a few selected particles.}
	\end{figure}

	In encounters with active grid sites, the particles' kinetic energy  changes by the amount $\Delta W$ (the amount depends on the kind of scatterers considered, and in any case we here address only the case of systematic acceleration, where $\Delta W$ is always positive, see Sect.\ \ref{s:particle}). After the interaction with a scatterer, a particle again moves into a random direction, until the next encounter with a scatterer, or until it reaches one of the grid boundaries. The smallest possible free travel distance between scattering events is given by the grid size ($\ell$). The scattering events are considered to be instantaneous, so the timing is determined by the free travel times $\Delta t = s/v,$ where $v$ is the particle's velocity and $s$  the distance it travels in between subsequent scatterings.

\subsection{Systematic acceleration by random electric fields}\label{ss:EF}
\subsubsection{Open boundary conditions}\label{sss:EFo}
A 2D version of this model was presented by \cite{Vlahos16}, which we extend here to a 3D acceleration volume. Following Eq.\ \eqref{e:dW_ef}, each scattering at an active grid point increases the particle's energy by
$\Delta W = |q| \left(\frac{V_A}{c}\; \delta B\right) \ell_{\rm eff}.$

The parameters used in this article are related to the plasma parameters in the low solar corona. We choose the strength of the magnetic field to be $B = \SI{100}{G}$, the density of the plasma $n_0 = \SI{e9}{cm^{-3}}$, the ambient temperature around \SI{10}{eV}, and the length $L$ of the simulation box is \SI{e9}{cm}. The Alfv\'en speed is $V_A \approx \SI{7e8}{cm/sec}$, so $V_A$ is comparable with the thermal speed of the electrons. We assume that $\delta B$ takes random values following a power-law distribution with index \num{5/3} (Kolmogorov spectrum), and $\delta B \in [\SI{.e-5}{G}, \SI{100}{G}]$. 
The effective electric field $E_{\rm eff} = (V_A/c) \delta B$ lies approximately in $E_{\rm eff} \in [2\cdot 10^{-7},2]\,$statV/cm.
We also assume the effective length $\ell_{\rm eff}$ to be a linear function of $E_{\rm eff}$, $\ell_{\rm eff} = a E_{\rm eff} + b$, where the constants $a$, $b$ are determined by restricting the size of $\ell_{\rm eff}$ to the range from \SI{10}{m} to \SI{1}{km} (see also \cite{Zhdankin13} for the statistical properties of UCS). 
Applying these values to Eq.~\eqref{e:dW_ef}, we find that the energy change $\Delta W$ varies between 
$10^{-13}\,$eV and  $10^8\,$eV
(due to the power-law distribution of $\delta B$, small energy changes are much more frequent). 

We consider the grid to be open, so particles can escape from the acceleration region when they reach any boundary of the grid. We assume in this set-up that only $R = \SI{10}{\percent}$ of the $N^3 = 601^3$ grid sites are active. The mean free path is thus given as $\lambda_{\rm sc} = \ell/R \approx \SI{1.7e7}{cm}$, which coincides with the value estimated numerically by tracing particles inside the simulation box.

As Fig.~\ref{f2a} shows, the particles execute a classical random walk on the grid in position space, in energy space though 
the dynamics is of the form of a systematic random walk,
see Fig.~\ref{f2b}, there are exclusively positive energy increments $dW$ 
(for details of the energy gain within the UCS see \cite{Guo15, Dahlin15, Matsumoto15,Isliker17}).

\begin{figure}[ht]
	\sidesubfloat[]{
	\includegraphics[width=0.45\columnwidth]{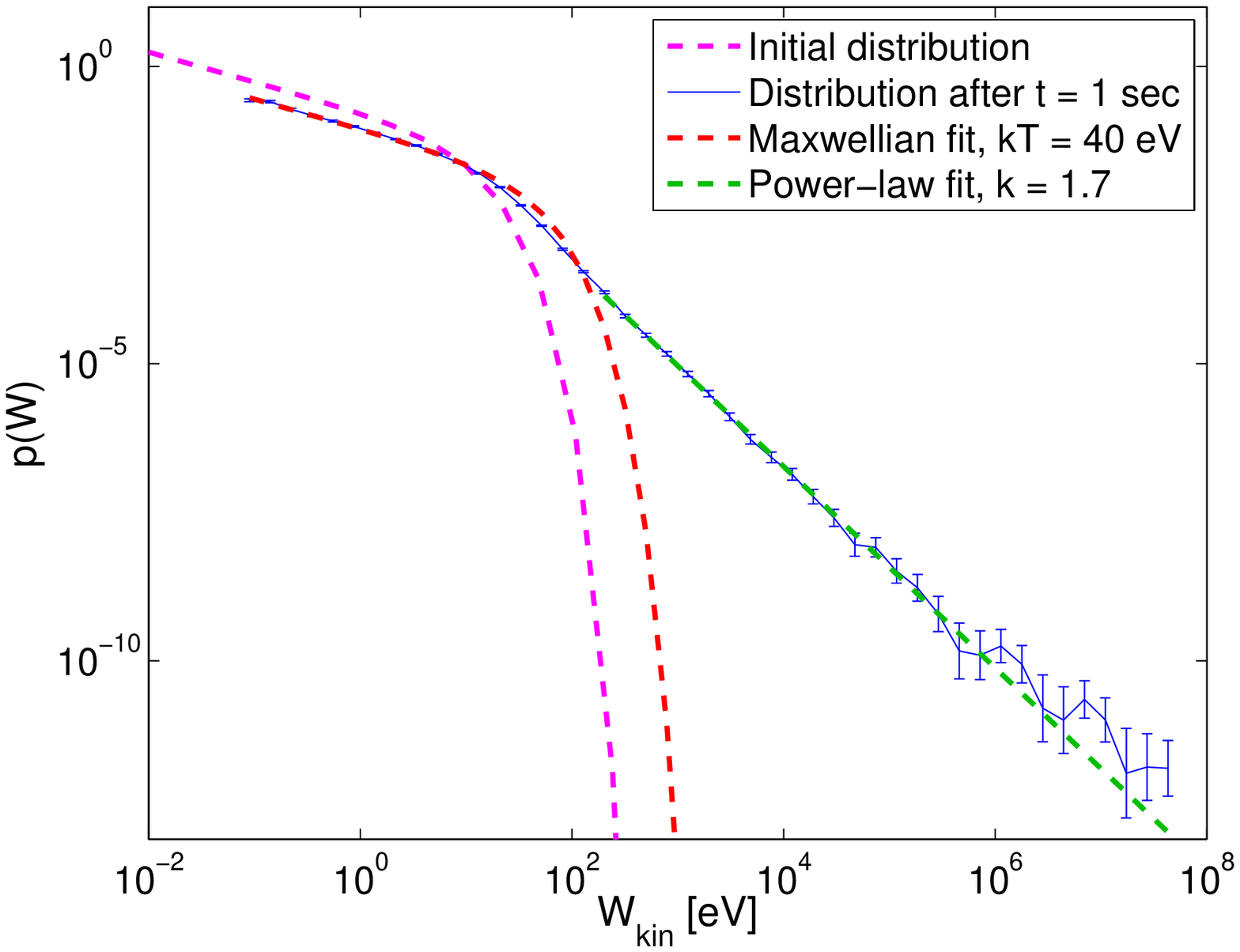}
        \label{f:EFo:nW}}
	\sidesubfloat[]{
	\includegraphics[width=0.45\columnwidth]{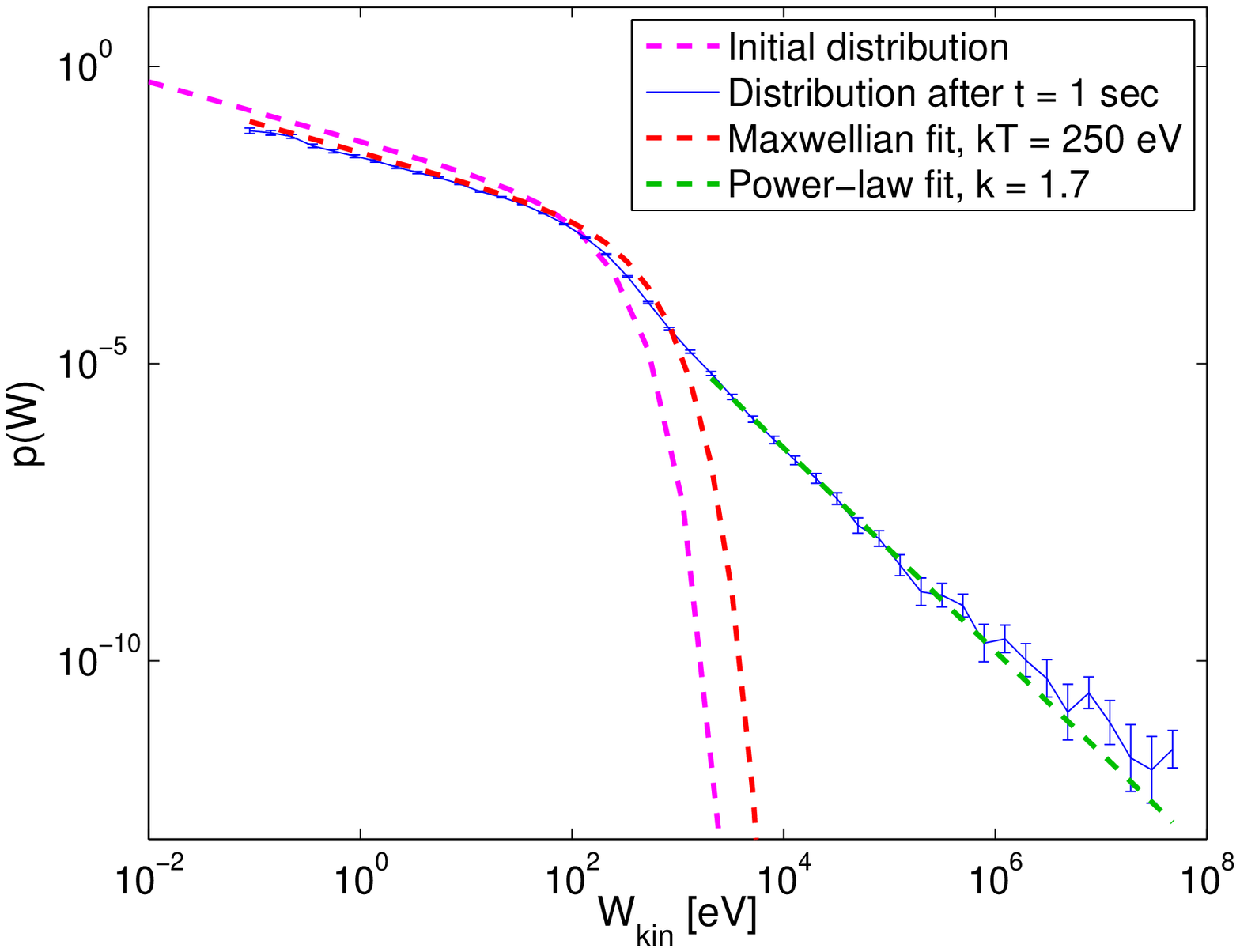}
        \label{f:EFo:nW:100}}%
    \caption{
    	Energy distribution at $t = 0$ and $t = \SI{1}{sec}$ of the electrons that have remained inside the box, for 
 	    an initial temperature of \protect\subref{f:EFo:nW}  $10\,$eV and \protect\subref{f:EFo:nW:100} $100\,$eV. 
} \label{f:EFo}
\end{figure}

We monitor the electron population, injected at $t = 0$, until half of it has escaped, which happens at $t_{1/2} = \SI{4.5}{sec}$. The electrons that remain inside the box at $t=1\,$sec are distributed as shown in Fig.~\ref{f:EFo:nW}.
The energization process heats the low energy particles below \SI{1}{keV}, where the distribution follows a Maxwellian with temperature \SI{40}{eV}. The high energy part of the distribution exhibits a clear power-law tail, with index $k \approx 1.7$, which extends from about \SI{1}{keV} to {100}{MeV}. The power-law tail is formed in a few milliseconds and it persists even when more electrons have escaped form the acceleration volume. This implies that the acceleration process is extremely fast.

	Using a higher initial temperature of $ 100\, $eV for the particles does basically not affect the high energy part of the final energy distribution, as Fig.~\ref{f:EFo:nW:100} shows, the extent and the index of the power-law tail remain almost unchanged,
	and only at the low energy part the particles are heated to a larger temperature of $250\,$eV.

\begin{figure}[ht]
	\sidesubfloat[]{\includegraphics[width=0.45\columnwidth]{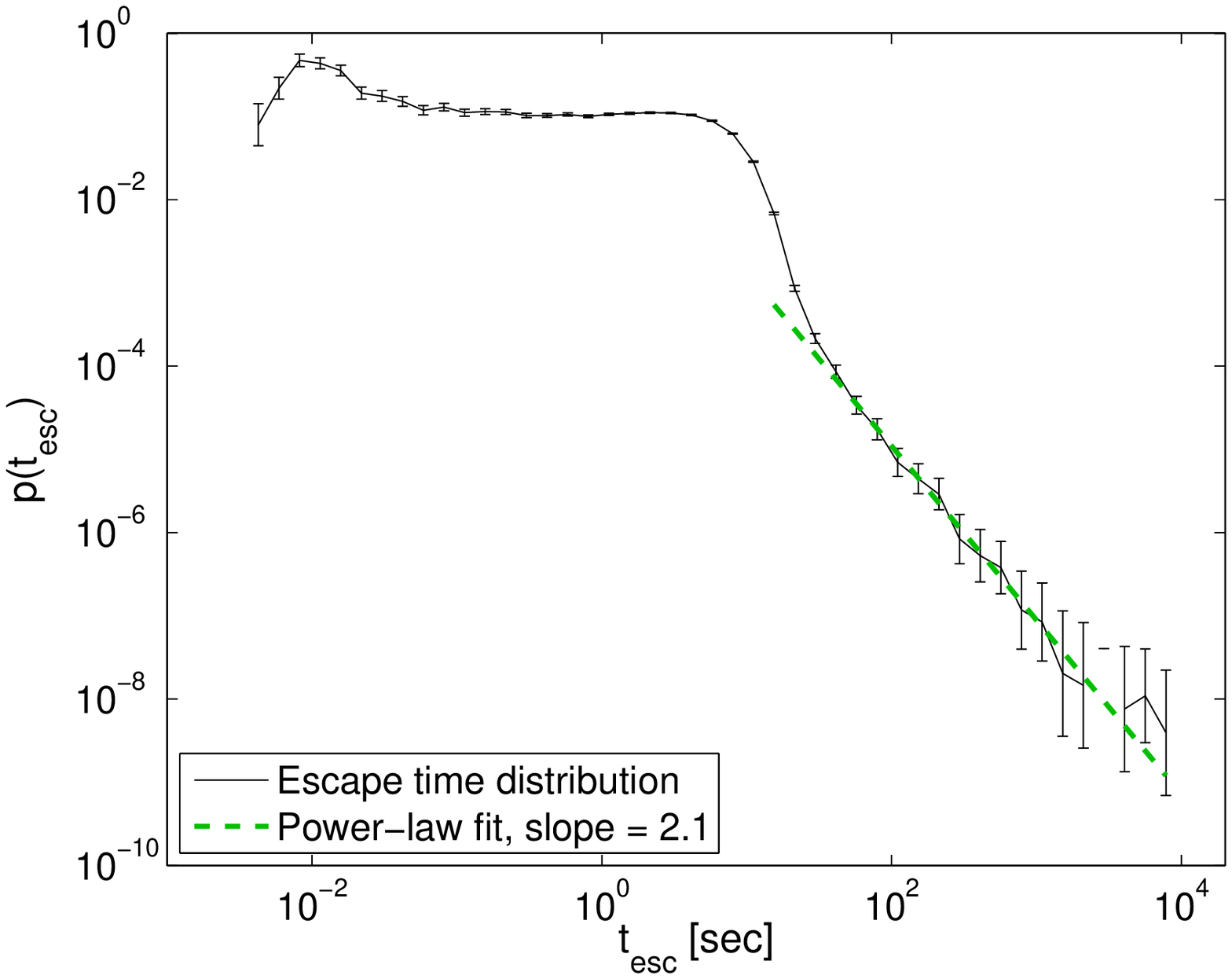}%
        \label{f:EFesc:nt}}\hfill
	\sidesubfloat[]{\includegraphics[width=0.45\columnwidth]{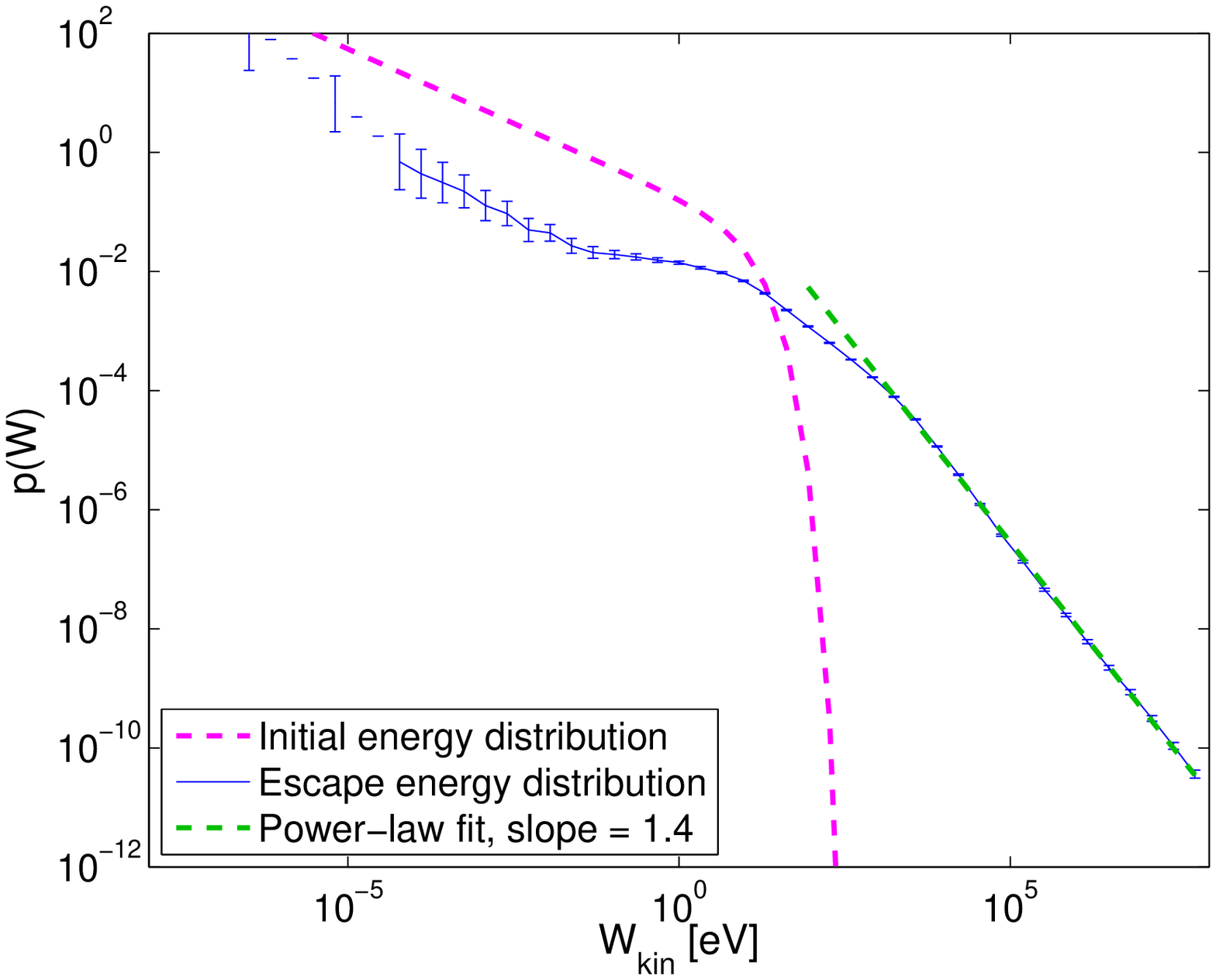}%
        \label{f:EFesc:nW}}\\
    \sidesubfloat[]{\includegraphics[width=0.45\columnwidth]{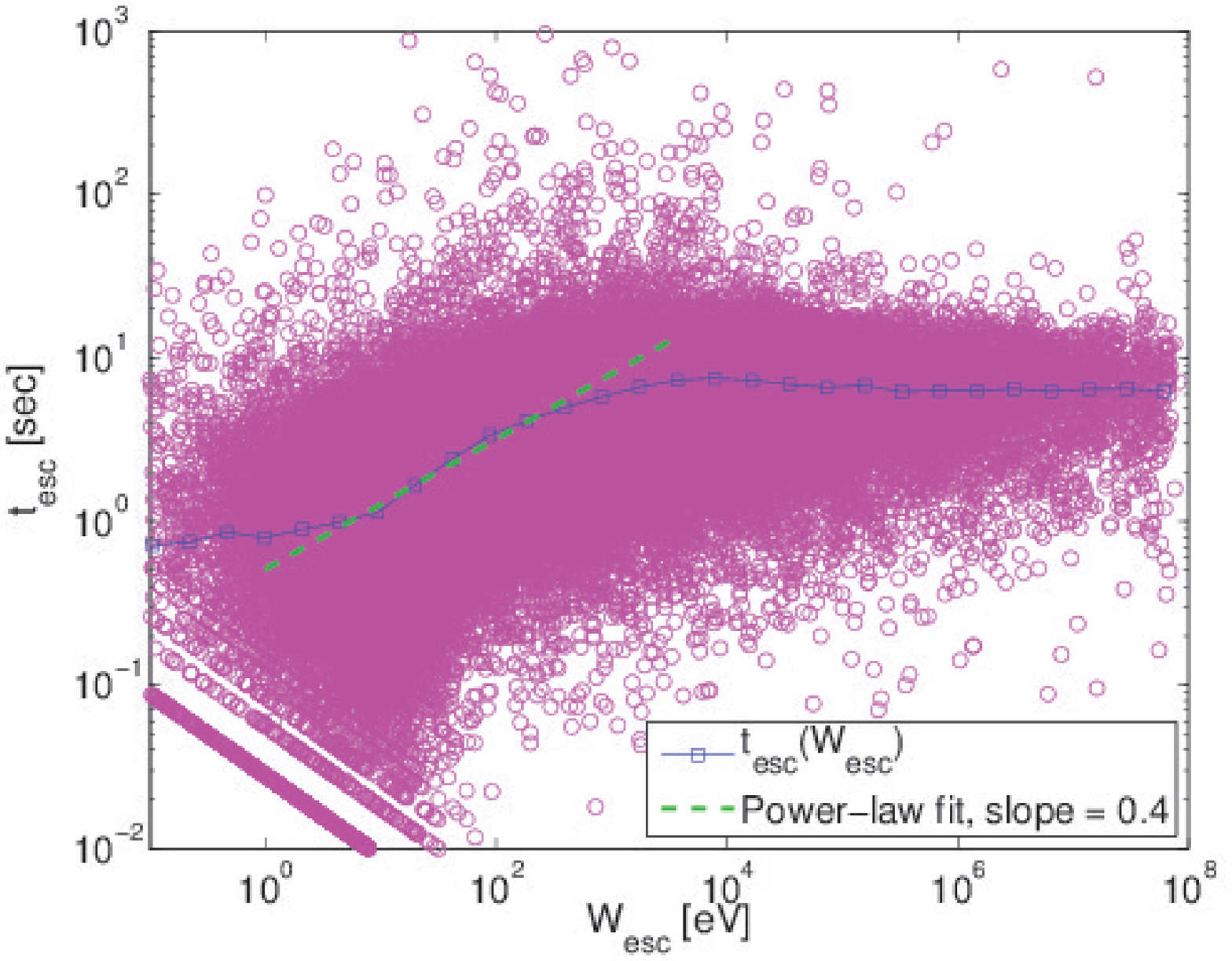}%
        \label{f:EFesc:tW}}\hfill
	\sidesubfloat[]{\includegraphics[width=0.45\columnwidth]{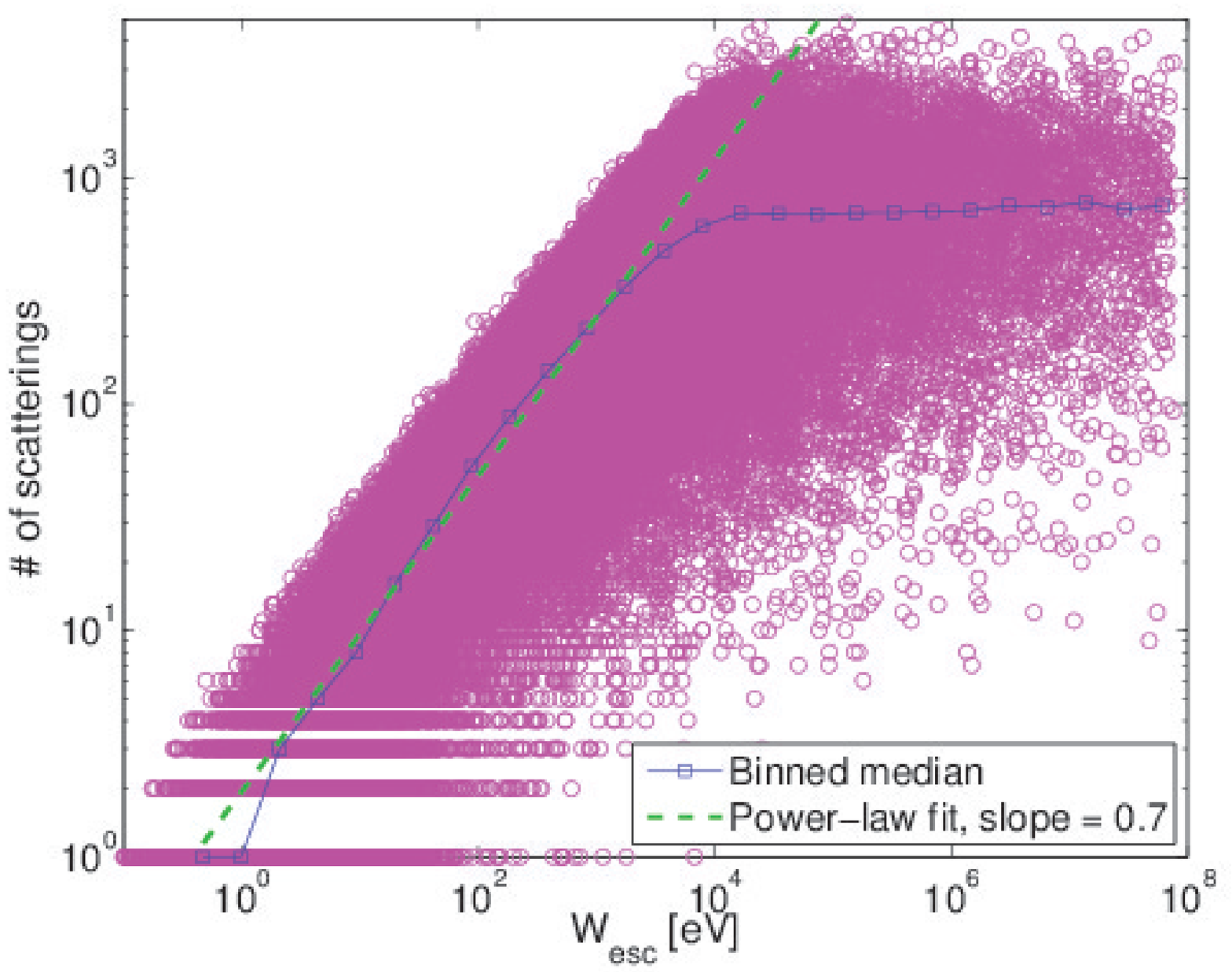}%
        \label{f:EFesc:kicksW}}%
    \caption{\protect\subref{f:EFesc:nt} The distribution of the escape times of the electrons. \protect\subref{f:EFesc:nW} The energy distribution of the electrons when they have escaped from the box. \protect\subref{f:EFesc:tW} The escape time as a function of the escape energy of the electrons. \protect\subref{f:EFesc:kicksW} The number of scatterings per particle as a function of the escape energy.
    }\label{f:EFesc}
\end{figure}

Each of the escaping electrons 
leaves with a different energy and at a different time, depending on its initial energy and the energization resulting from the scattering events. We term the time a particle reaches any boundary of the acceleration volume its escape time $ t_{\rm esc} $, and the energy it carries at that moment its escape energy. The distribution of the escape energy, presented in Fig.~\ref{f:EFesc:nW}, exhibits a shape very similar to the one of the particles that stay inside (Fig.~\ref{f:EFo:nW}), a power-law tail is formed with index 1.4 at energies 
above \SI{1}{keV}.
Fig.~\ref{f:EFesc:nt} shows the distribution of the escape times for all the escaping electrons; it is uniform up to about \SI{10}{sec}, and then it turns into a power-law tail with index $\approx 2$, up to the time of \SI{.e4}{sec}. The median value of the escape times of all electrons can be used as an estimate for the characteristic escape time from the system, $t_{\rm esc} \approx \SI{4.5}{sec}$, which coincides with its half time $t_{1/2}$ defined above. By using binned statistics, the escape time can be interpreted to be a function of the escape energy, as shown in Fig.~\ref{f:EFesc:tW}. We observe two distinct regions, the high energy one, extending from \SI{1}{keV} to \SI{1}{MeV}, and the low energy one, below \SI{1}{keV}. At low energies, there is a power-law functional form, $t_{\rm esc} \propto W_{\rm esc}^{0.4}$. 
At higher energies, the function becomes constant, assuming the value
$t_{\rm esc} \approx 5$. 
This indicates that the high energy particles, which form the power-law tail of the energy distribution, 
stay longest in the acceleration volume, and with that,
they basically determine the mean escape time. Also, the degree of energization is thus directly correlated with the time that the particles stay inside the system only for low energy particles, the high energy particles presuppose the largest, yet constant and independent of the particles' energy, time to be accelerated in the system. 
The number of scatterings as a function of the final energy (again estimated by  using binned statistics) yields a very similar picture, see Fig.\ \ref{f:EFesc:kicksW}, only for the low energy particles the final energy increases with increasing number of scatterings, the high energy particles all undergo approximately the same number of 1000 scatterings, independent of their final energy.

{There are two key parameters for the turbulent reconnection analyzed here. } The first is the mean free path in-between encounters with the scatterers, $\lambda_{\rm sc}$, which is controlled by the density of the latter, $n_{\rm sc}$, as defined in Sec.~\ref{ss:setup}. Changing any of the parameters while keeping the scatterer density fixed has no effect on the results. We thus for simplicity fix both the size of the acceleration volume and the number of grid-points to constant values, so that the density of the scatterers depends solely on the active-point ratio $R$. As this ratio varies in the range $0.05 \leqslant R \leqslant 0.2$ (i.e. $\SI{e7}{cm} \leqslant \lambda_{\rm sc} \leqslant \SI{e8}{cm}$), the acceleration time remains extremely small, while the escape time varies from \SI{5.5}{sec} to \SI{1.7}{sec}, with an almost linear decrease. The energy distribution, on the other hand, maintains its power-law form. The second important free parameter of the system is the index $z_{\delta B}$ of the power-law distribution of the fluctuations of the magnetic field, $\delta B$. As it varies from $1 \lesssim z_{\delta B} \lesssim 3$ the power-law index $k$ of the particles' energy distribution follows approximately the evolution of $ z_{\delta B} $. 

The results discussed so far refer to electrons. Using ions in the simulations, we find that the evolution of the energy distribution exhibits the same features as the one of the electrons, except for the time scales involved. The ion population half time amounts to \SI{195}{sec}, as does the median escape time (the escape times now vary from \SI{0.1}{sec} to \SI{.e6}{sec}), while the acceleration time is approximately \SI{250}{sec}. Keeping in mind these different time scales, the escape characteristics are also similar (see Fig.~\ref{f:EFesc}): the escape energy and escape time distributions have a power-law tail with indices \num{1.5} and \num{2.1}, respectively, and $t_{\rm esc}$ and number of scatterings per particle as functions of $W_{\rm esc}$ retain a power-law shape at low energies and turn over to become constant at large energies. In other words, ions follow very closely the characteristics of the electrons, yet with a remarkable delay of several minutes (for the parameters used here).

In oder to investigate the role of collisions, we apply a modified version of the collision model of \cite{Lenard58}, where
the charged particles undergo Coulomb collisions with a background plasma population of temperature $T$ (which we choose equal to the particles' initial temperature). 
As described in \cite{Pisokas16}, if a particle with initial velocity $v(t)$ travels the distance $s$ 
between two subsequent encounters with scatterers during a time interval $\tau$, then the final velocity $v(t+\tau)$ can be computed by an analytical expression \citep{Gillespie1996},
\begin{equation}\label{e:colls_sol}
    v(t+\tau) = v(t)\mu + \sigma_v N_1,
\end{equation}
where $\mu = e^{-\nu_v\tau}$, $\sigma_v^2 = \frac{k_B T}{m}\left(1-\mu^2\right)$, and $N_1$ is a Gaussian random variable with mean 0 and standard deviation 1. We note that in this modified model the collision frequency $ \nu_v $ is velocity dependent, $\nu_v \propto 1/v(t)^3$, as appropriate for a fully ionized plasma, and that $v$ here is the non-relativistic speed that is bounded by the speed of light.

\begin{figure}[ht]
    \includegraphics[width=0.65\columnwidth]{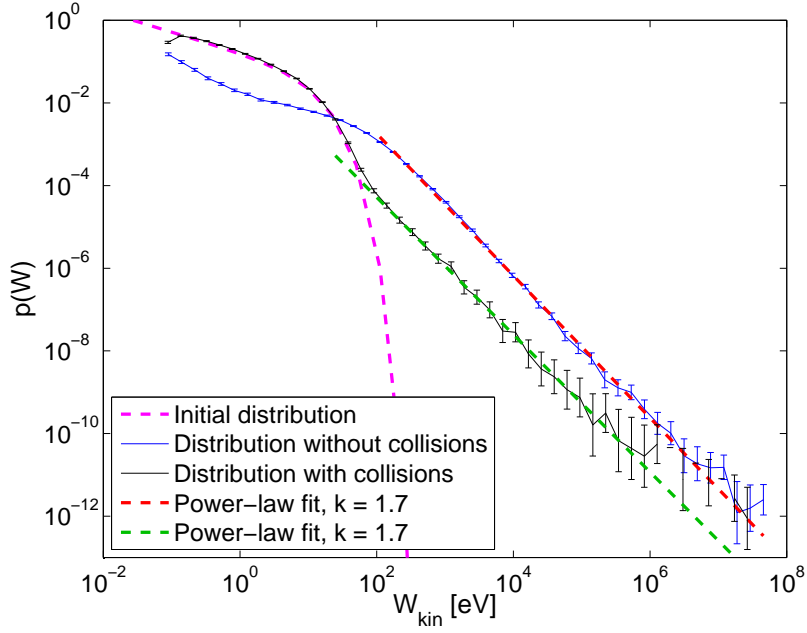}
    \caption{Energy distribution for electrons remaining inside the simulation box at $t = 0$ (magenta) and at $t = \SI{4.5}{sec}$ with collisions (black) and without (blue).}\label{f:EFoC_nW}
\end{figure}

In Fig.~\ref{f:EFoC_nW} we compare the energy distribution of the electrons remaining inside the acceleration volume for the cases with and without collisions at $t = t_{\rm acc} = \SI{3.5}{sec}$. The collisional mean free path is 180 times smaller than the mean free path between scatterings ($\lambda_{\rm coll} = \SI{9.27e4}{cm}$), so collisions are relevant, yet they affect only the low energy part of the distribution, below \SI{100}{eV}, the high energy part shows a power-law shape with the same index as in the collisionless case, $k = 1.7$

\subsubsection{Periodic boundary conditions}\label{sss:EFp}
We now 
impose periodic boundary conditions, i.e.\ any particle that reaches a boundary grid-site continues its motion from the corresponding grid-site on the opposite boundary, without any other changes in the set-up. 

\begin{figure}[ht]
	\includegraphics[width=0.65\columnwidth]{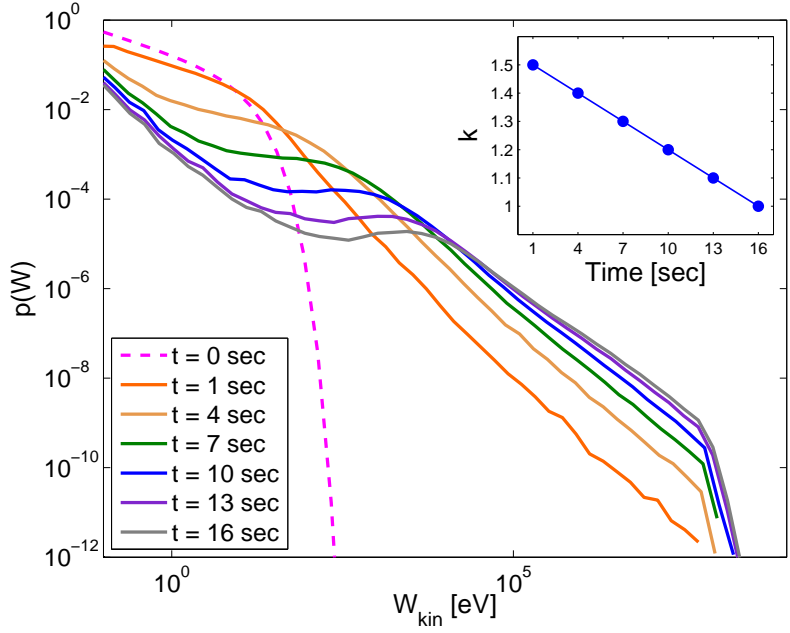}
    \caption{Evolution of the energy distribution until it reaches the asymptotic index $k = 1$. Inset: the evolution of the power-law index.}\label{f:EFp_nW}
\end{figure}

The evolution of the energy distribution of electrons  is shown in Fig.~\ref{f:EFp_nW}, together with the temporal evolution of the power-law exponent of its tail.
The energy distribution develops a power-law in a few milliseconds, with an index that ultimately drops to an asymptotic value of $k = 1$ at $t \approx \SI{16}{sec}$. After that time the distribution just extends to higher energies, without changing its power-law shape anymore. We note that this result agrees very well with numerous 3D PIC simulations \cite[see for example][]{Guo15, Matsumoto15}.

Again, the crucial parameter for the evolution of the system is the mean free path between scatterings $\lambda_{\rm sc}$. 
An increase of the ratio of active grid sites $R$ from \num{0.05} to \num{0.2} 
leads to a decrease of
the time for the power-law shaped energy distribution to reach a specific  index, namely from \SI{28}{sec} to \SI{7}{sec} for $k = 1.1$, and from \SI{33}{sec} to \SI{7}{sec} for $k = 1$.  

As for $z_{\delta B}$, we observe that  

for indices $z_{\delta B}$ below 2 the asymptotic energy distribution exhibits a power-law with $k = 1$, and for $2 \lesssim z_{\delta B} \lesssim 3$ the index of the asymptotic power-law is $k \geq z_{\delta B}$.


\subsection{Systematic Fermi acceleration at contracting islands}\label{ss:SF}

We now consider systematic Fermi acceleration, with 
the energy gain during a particle-scatterer interaction given by 
Eq.\ (\ref{e:dW_sf}), and otherwise maintaining the setup as described in the first part of the current section. Using the parameters presented in Sect.~\ref{ss:EF}, we find that the energization is much more effective, with a typical energy gain of $\approx\SI{600}{eV}$ per scattering event. We thus repeat the analysis in a larger acceleration volume with characteristic length $L = \SI{5e9}{cm}$, with open boundaries, and with only 5\% of the grid points being active, which leads to a larger mean free path among the scatterers, $\lambda_{\rm sc} = \SI{1.67e8}{cm}$. 
With these changes, the energy distribution of the electrons exhibits a narrow power-law distribution with index $k \approx 3.5$ at $\SI{100}{msec}$, which, with increasing time, gains in extent and the power-law index  decreases, until it stabilizes at the value of $k =1$ at \SI{1}{sec}; for larger times, the power-law still widens up, yet the index does not change anymore.

\begin{figure}[ht]
	\sidesubfloat[]{\includegraphics[width=0.45\columnwidth]{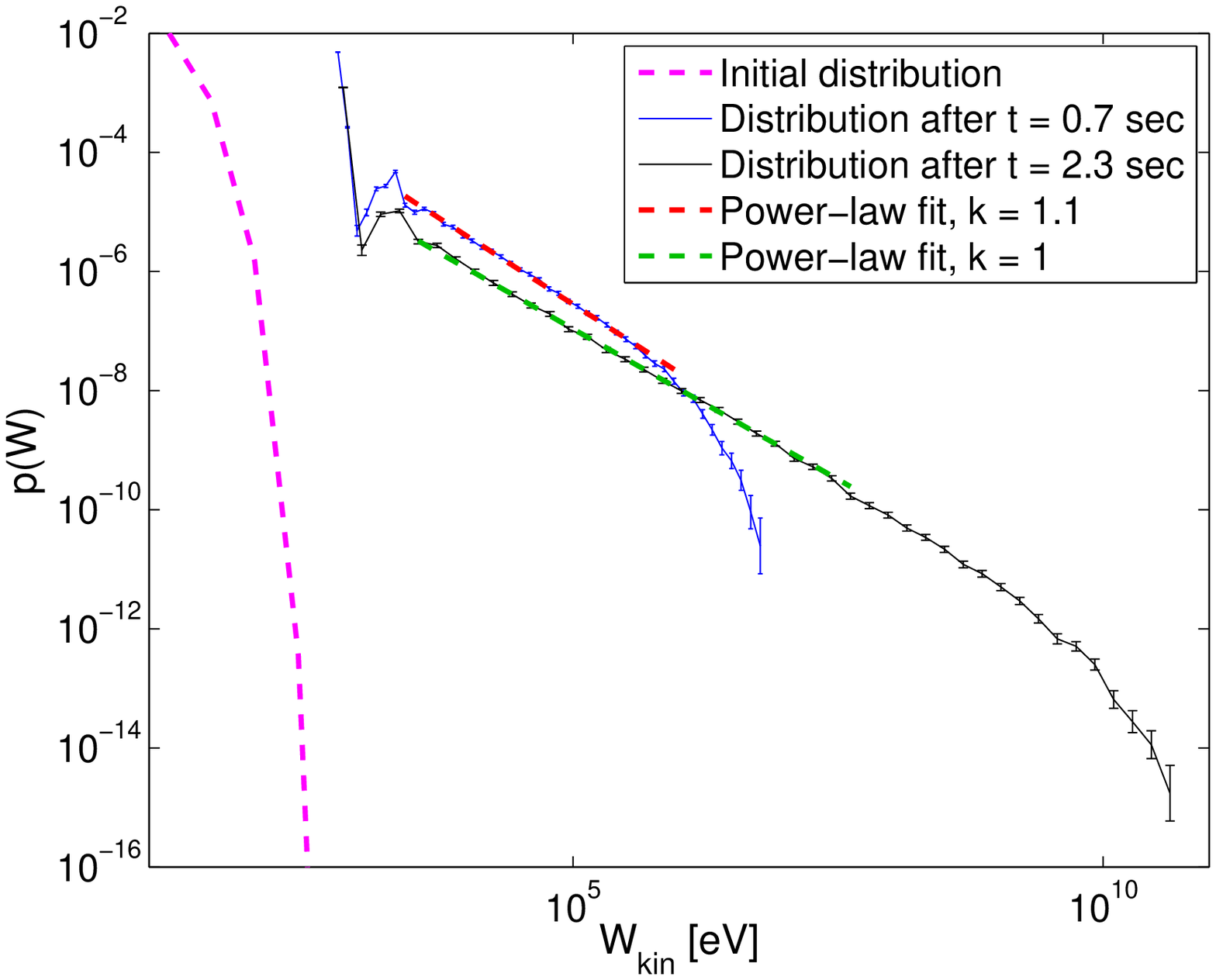}%
        \label{f:SFo:nW}}\hfill
    \sidesubfloat[]{\includegraphics[width=0.45\columnwidth]{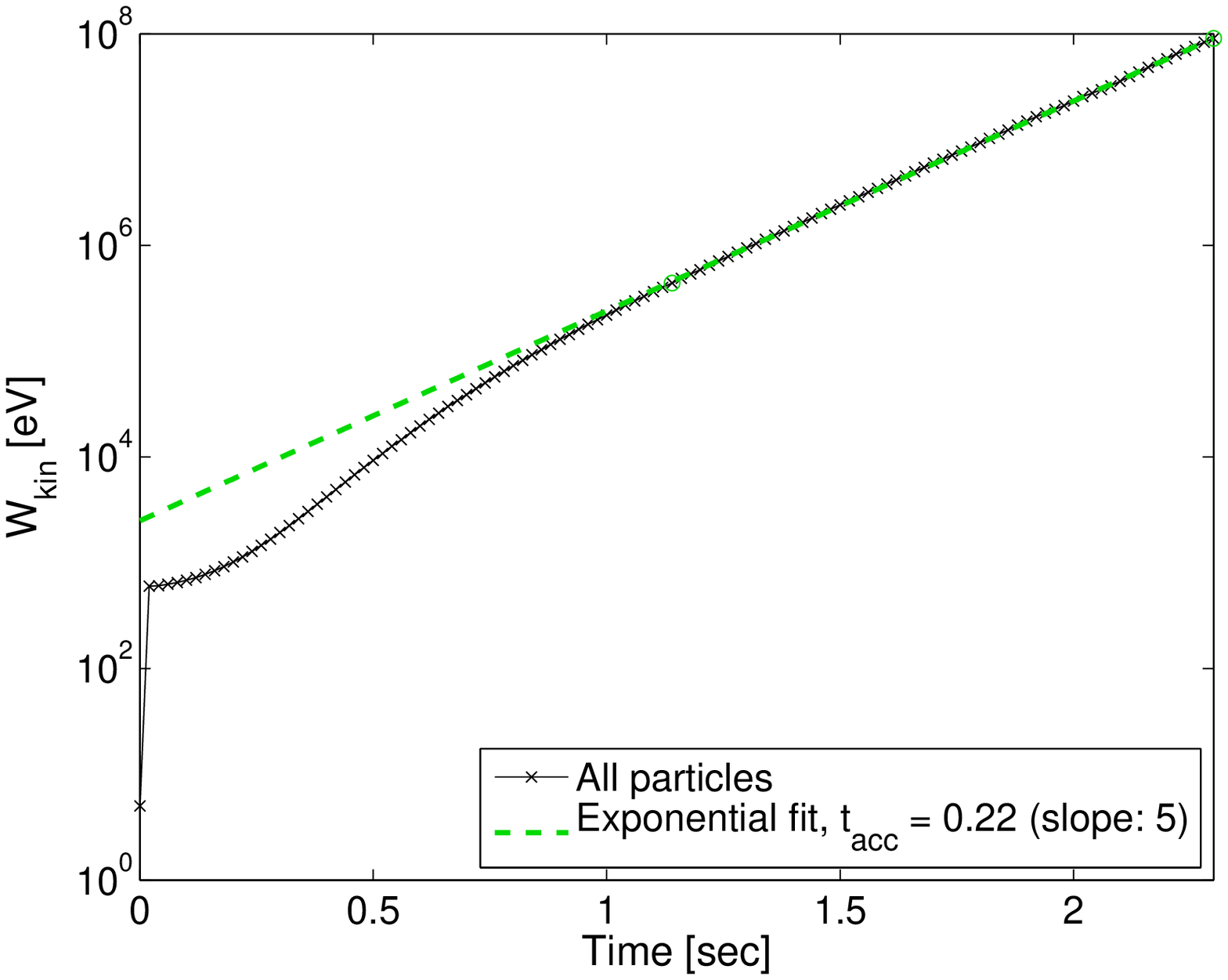}%
        \label{f:SFo:mW}}%
    \caption{\protect\subref{f:SFo:nW} Energy distribution at $t = 0$ (magenta), $t = \SI{0.7}{sec}$ (blue), and  $t = t_{1/2} = \SI{2.3}{sec}$ (black) for the electrons, together with the corresponding power-law fits, $k = 1.1$ (red), and $k = 1$ (green). \protect\subref{f:SFo:mW} Mean kinetic energy of the electrons as a function of time (black) with an exponential fit (green).
   (b)  }\label{f:SFo}
\end{figure}

We present two time instances of the evolution of the energy distribution in Fig.~\ref{f:SFo:nW}. The first one is the energy distribution at $t = \SI{.7}{sec}$, where the power law index has attained the value $k = 1.1$.
At this time, close to the asymptotic state, almost 90\% of the particles are still inside the system, the high intensity of the energization process leads the system to its asymptotic state before a considerable percentage of the initial population escapes. The application of periodic boundary conditions
results therefore in similar power-law distributions, and it has only  minor effects on the characteristics of the system before its stabilization. The other time-instance of interest is the half time of the system, when 50\% of the particles have escaped from the acceleration box, which once again coincides with the escape time, defined as the median of the individual escape times of the particles, $t_{1/2} \approx t_{\rm esc} \approx \SI{2.3}{sec}$. The power-law shaped energy distribution still exhibits the index $k = 1$, which persists for longer times, even when the vast majority of the particles has escaped. We can clearly conclude that the power law shape and its index are characteristic for the energization process and they are independent of the boundary conditions.

By solving Eq.~(\ref{e:dW_sf}) one would expect an exponential growth of the mean energy \cite[see][and Eqs.\ (2) and (3) therein]{Pisokas16}. The mean energy of the electrons as a function of time in Fig.~\ref{f:SFo:mW} clearly agrees with this prediction, from which we can estimate the acceleration time for the system as the growth rate, $t_{\rm acc} = \SI{0.22}{sec}$.

Also for systematic Fermi acceleration at contracting islands, the key parameter of the system remains the mean free path of the particles, $\lambda_{sc}$, and it actually is reciprocally related to the acceleration and stabilization times, yet it has no significant effect on the escape time, which can be attributed to the high intensity of the energization process. 



\section{Transport properties of particles inside a turbulently reconnecting plasma}\label{s:transport}

For one case of electric field acceleration and one case of 
acceleration at contracting islands, we now determine the 
transport coefficients $ F(W) $, Eq.\ (\ref{eq:FW}), and $ D(W) $, Eq.\ (\ref{eq:DWW}), as a function of energy. We then insert the transport coefficients into the FP equation, Eq.\ (\ref{diff}) (including the escape term, and with $ Q=0 $), and solve the latter numerically, as described in Sect.\ \ref{ss:FPtheor}, which either, as shown in \cite{Vlahos16}, can be successful in reproducing the energy distribution of the test-particles, or, as shown in \cite{Isliker17}, can fail to do so, and in this case, through an analysis of the energy increments [Eq.\ (\ref{delta_w})], it can become obvious that only an FTE approach is appropriate. In the case of failure of the FPE, we determine the parameters of the FTE, Eq.\ (\ref{fracth}), and solve it numerically, in the way described in Section \ref{ss:FTEtheor}.

\subsection{Electric field acceleration}

\subsubsection{Fokker Planck approach}

We first consider acceleration by the electric field, and more specifically  the case presented in Fig.\ \ref{f:EFo:nW}, for final time $1\,$s,
with open boundaries, and without collisions.
In Fig. \ref{f:EFesc:FD}, the diffusion and convection coefficients at
$ t = 1\, $s, as functions of the energy, are presented. There is a high level of noise in the estimate, which is a first indication that the FP approach might be problematic. Although there is no obvious functional form in the data, we performed a power law fit to them in the higher energy part above $ 1\, $keV, 
which yields the power-law indexes $a_D = 0$ and $a_F = 0.59$. For energies below $1\ keV$, we interpret the data as the two transport coefficients being constant.
In order to verify the estimates of the transport coefficients,
we insert them in the form of the fit into the FP equation (Eq.\ (\ref{diff})) and
solve the FP equation numerically. 
The resulting energy distribution 
is shown in Fig.~\ref{f:EFesc:FP}, together with the energy distribution from the lattice gas model. There obviously is a large discrepancy between the FP solution and the particles' energy distribution, the FP solution is basically flat in the entire energy range, whereas the particles' distribution shows a decaying and extended power-law tail.

One might attribute the failure of the FP approach to the choice of fit we made to the transport coefficient data in Fig.~\ref{f:EFesc:FD} and try to improve it. Yet, there is an inherent problem in the estimate of the transport coefficients that becomes obvious when considering the distribution of energy increments $ p(w) $ that we show in Fig.~\ref{f:EFesc:p_w}. 
This distribution exhibits an extended power-law shape, with index $-1.56$. From the definition of the energy increments in eq.~\eqref{delta_w} it is clear that the transport coefficients $ F(W) $ and $ D(W) $ [eq.~\eqref{eq:FW} and eq.~\eqref{eq:DWW}] basically are the first and second moment of the distribution of energy increments, and thus they theoretically are infinite, and in practice, when estimated with binned statistics, they are dominated by statistical noise. In other words, in the case at hand, the concept of transport coefficients in the classical sense is ill-defined.

\subsubsection{Fractional transport equation approach}

\begin{figure}[ht]
    \sidesubfloat[]{\includegraphics[width=0.41\columnwidth]{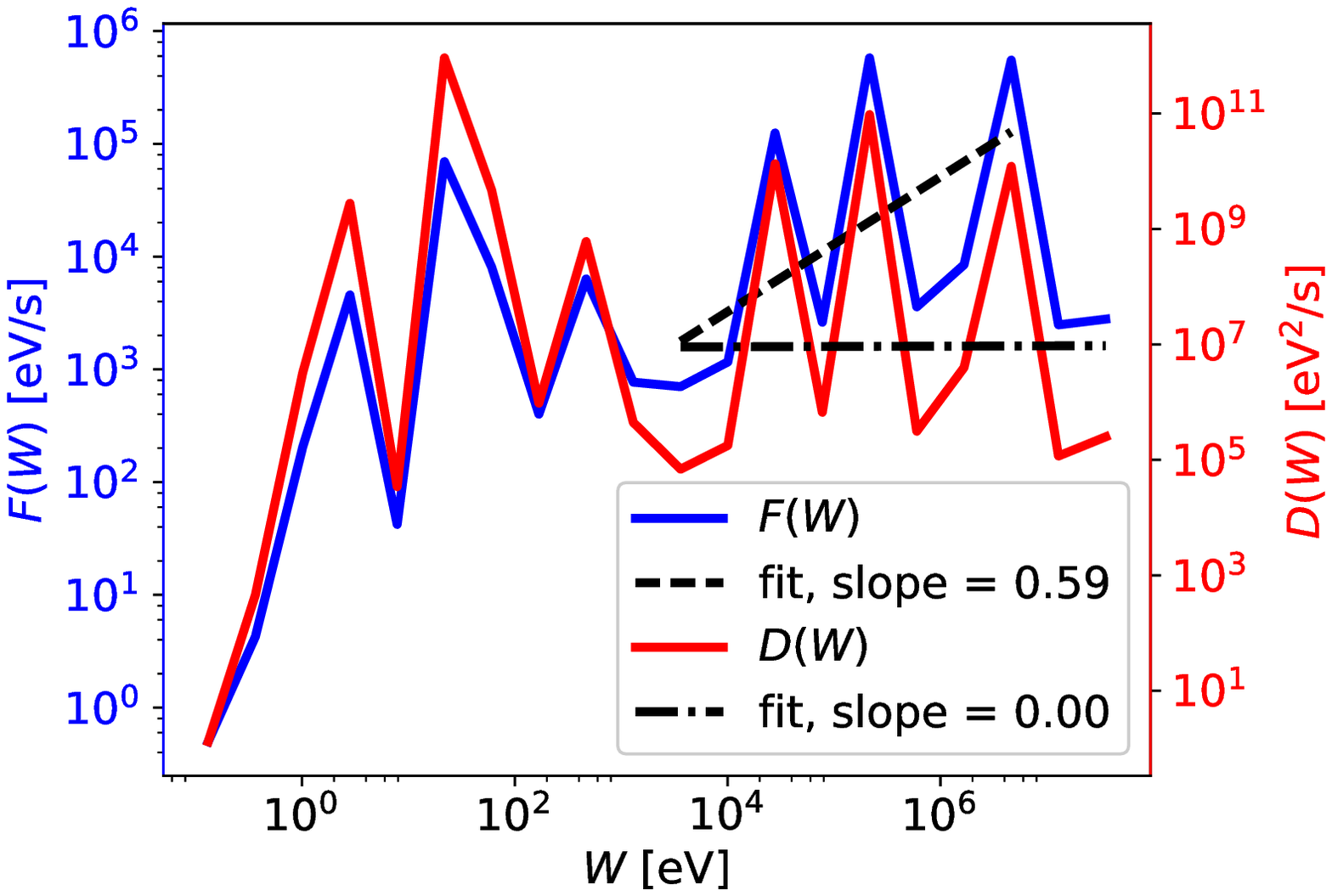}%
        \label{f:EFesc:FD}}\hfill
	\sidesubfloat[]{\includegraphics[width=0.41\columnwidth]{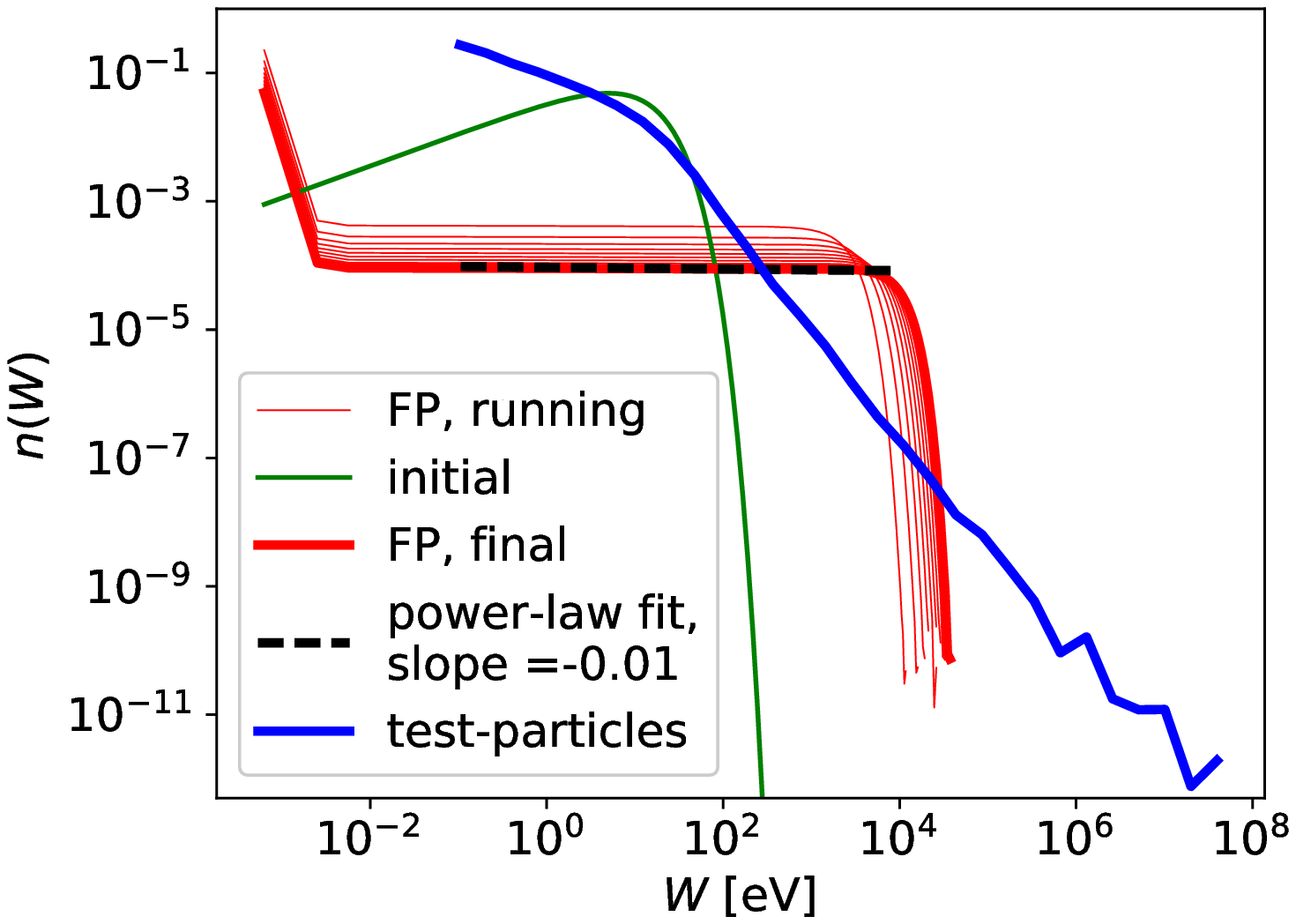}%
        \label{f:EFesc:FP}}\\
	\sidesubfloat[]{\includegraphics[width=0.41\columnwidth]{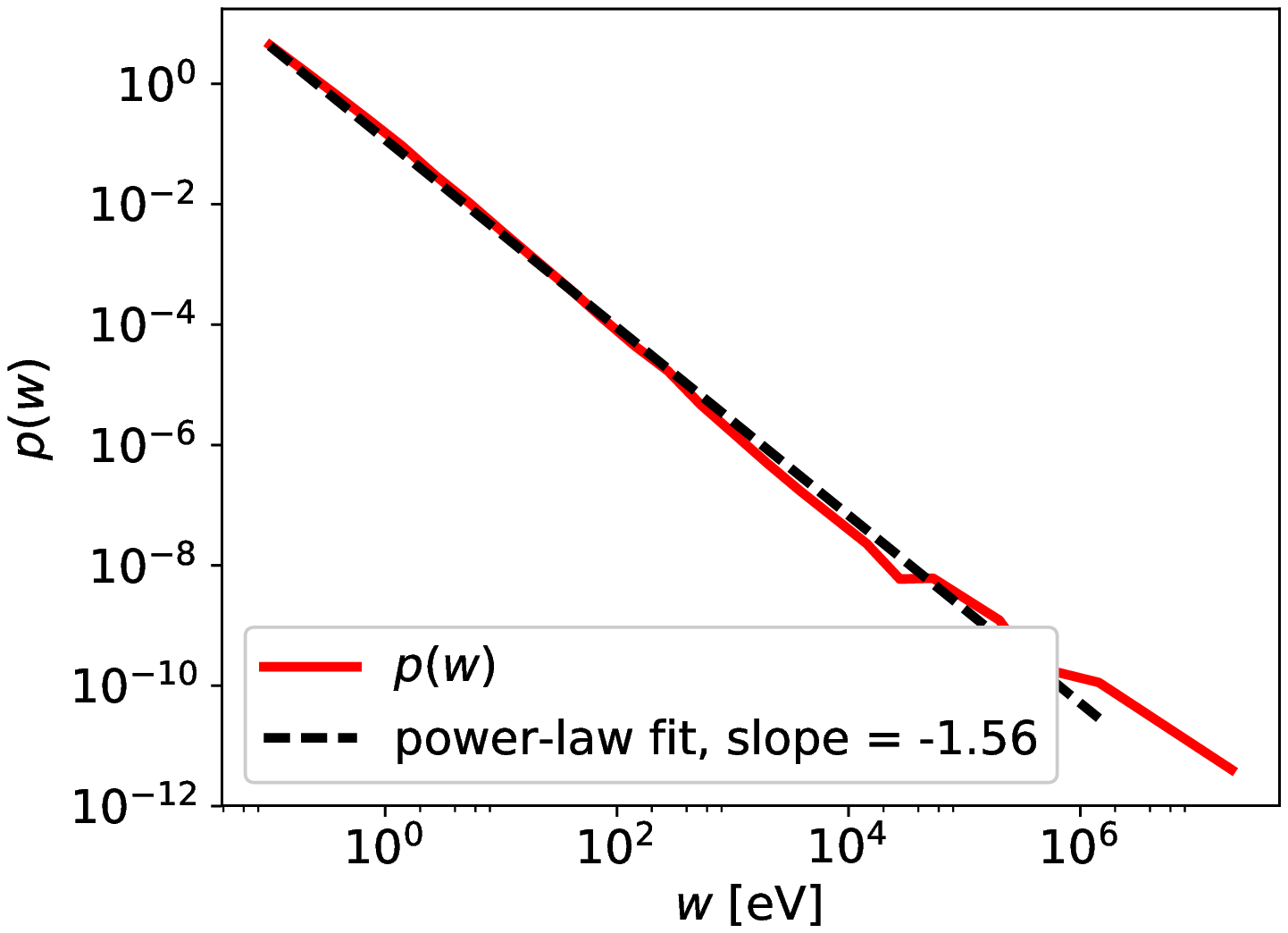}%
        \label{f:EFesc:p_w}}\hfill
	\sidesubfloat[]{\includegraphics[width=0.41\columnwidth]{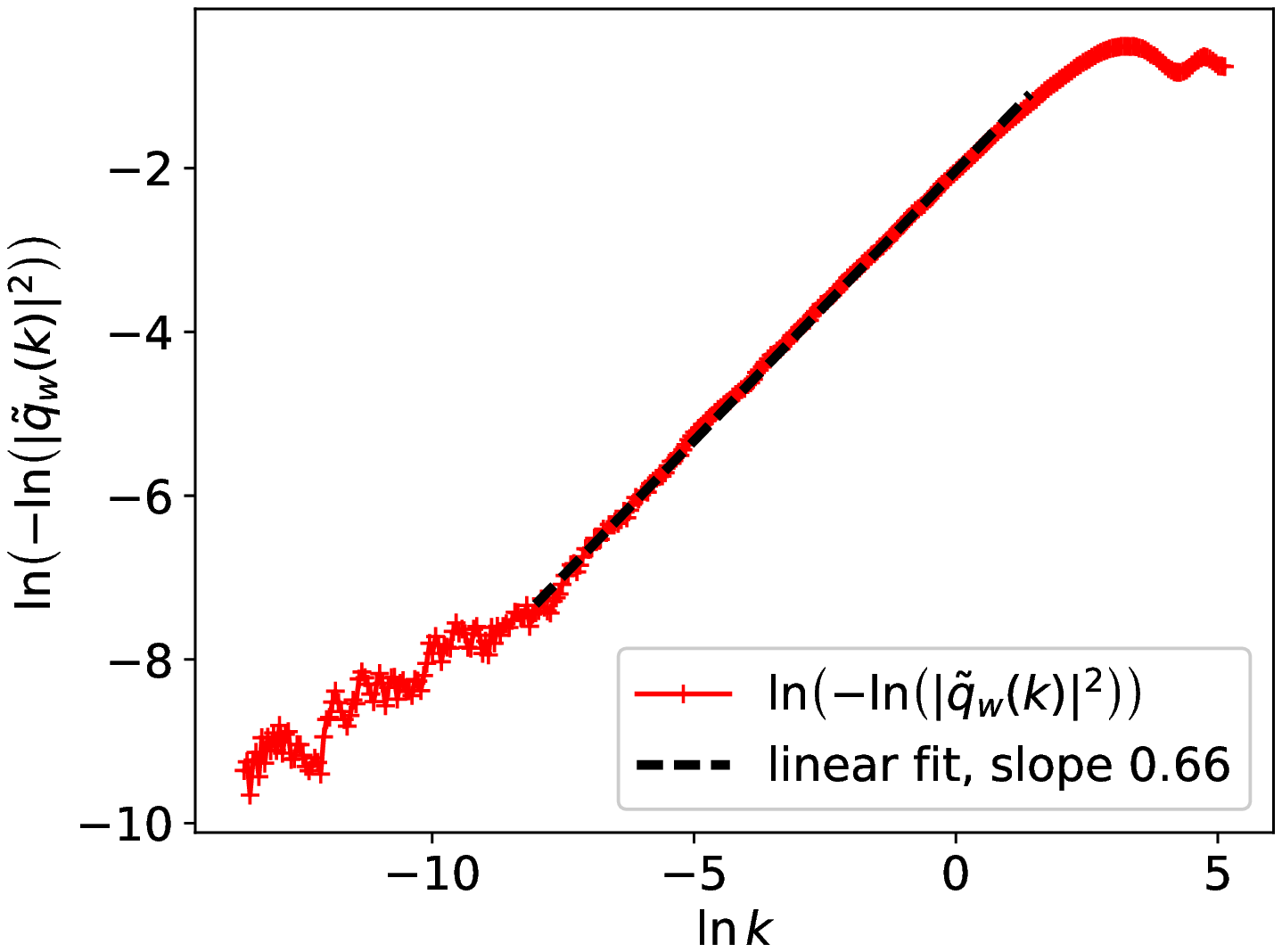}%
        \label{f:EFesc:char_fun}}\\
	\sidesubfloat[]{\includegraphics[width=0.45\columnwidth]{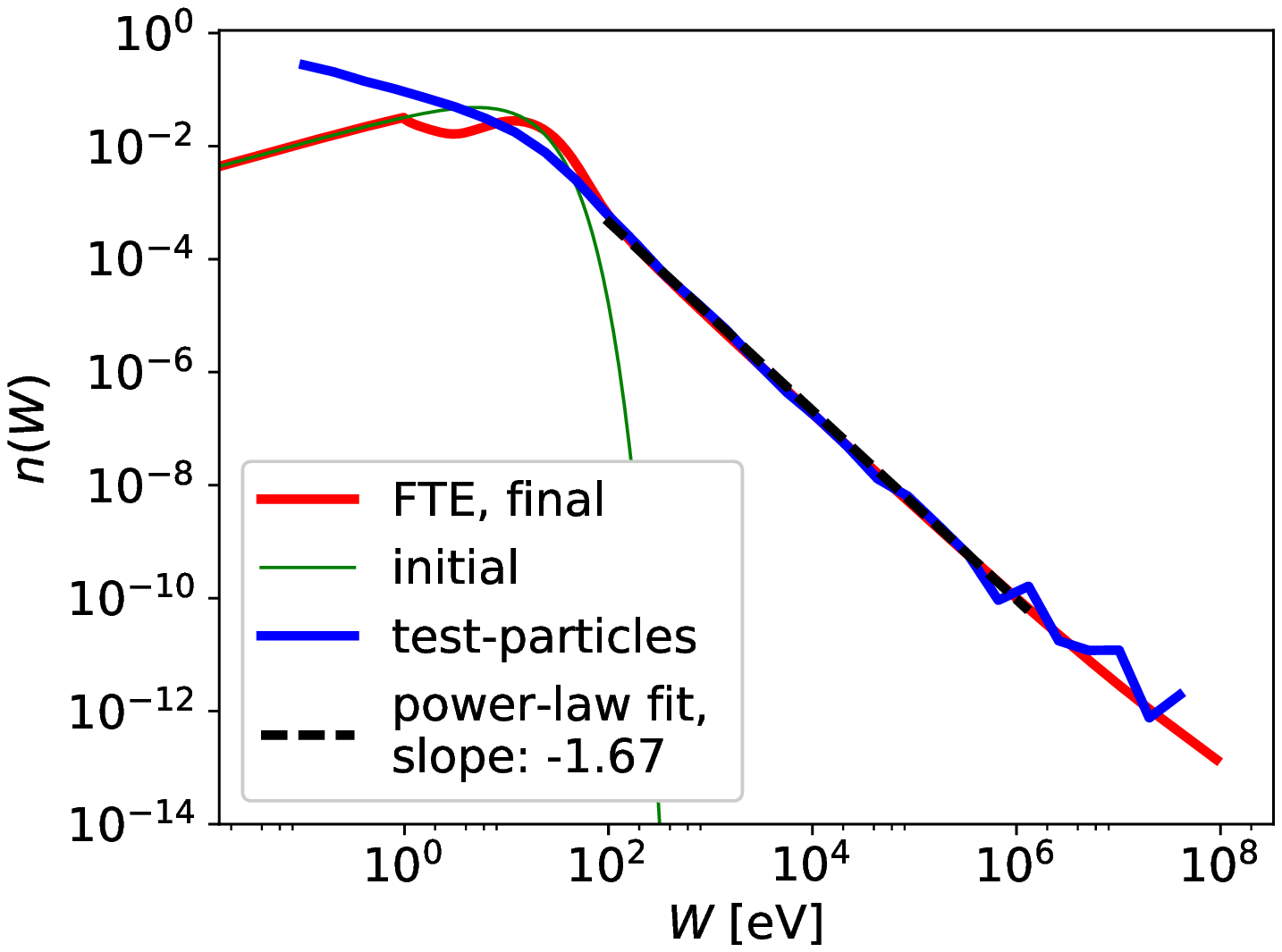}%
        \label{f:EFesc:FTE}}%
    \caption{
    \textit{Acceleration by electric fields:}
    \protect\subref{f:EFesc:FD} Convective $F$ and diffusive $D$ transport coefficients as a function of $W$ at $t=1\,$s; 
    \protect\subref{f:EFesc:FP} solution of the FP equation, together with the energy distribution from the lattice gas simulation; \protect\subref{f:EFesc:p_w} the distribution $p(w)$ of energy increments $w$; 
    \protect\subref{f:EFesc:char_fun}; the estimator of the characteristic function; 
    \protect\subref{f:EFesc:FTE} solution of the FTE, together with the energy distribution from the lattice gas simulation.}\label{f:EFesc:transport}
\end{figure}

Given that the distribution of energy increments $ p(w) $ has a power-law tail as do have the stable Levy distributions, we proceed to the FTE as a transport model to reproduce the particles' energy distribution.
The order of the fractional derivative $\alpha$ can be inferred from the index $z$ of the power-law tail of $p_w(w)$
in Fig.~\ref{f:EFesc:p_w} as $\alpha = -z-1 = 0.56$.
On the other hand, the 
characteristic function method [see Section \ref{ss:FTEtheor} and eq.~\eqref{e:char_fun}] yields $\alpha = 0.66$ and  
$a = 0.066$ for the scale parameter. 
Fig.~\ref{f:EFesc:FTE} shows the numerical solution of the FTE at time $t=1\,$s, using $\alpha$ as estimated through the characteristic function method, together with the particles' energy distribution. The FTE successfully reproduces the power-law tail of the particles' distribution in its entire extent.  

When changing the scale parameter $a$, 
we find that it does not affect the shape and power-law index of the 
FTE's solution, it just causes a shift of the solution in the vertical direction. We also find that the characteristic function method is not very precise in its estimate of the scale parameter $ a $, so that we rather consider the estimate of $ a $ as indicative and change it to fine-tune the coincidence of the FTE solution with the particle data in the vertical direction of the plot. For Fig.~\ref{f:EFesc:FTE} we finally have used a value of $ a = 3.33 $. After all, with the methods employed, we achieve a reasonable precision in the estimate of the order of the fractional derivative, and we get indicative values for the scale parameter, obviously though quantitatively more precise methods are desired and should be investigated, above all for the latter.


\subsection{Acceleration at contracting islands}

\subsubsection{Fokker Planck approach}

We now turn to acceleration at contracting islands, considering the case presented in Fig.~\ref{f:SFo:nW}, for final time $0.7\,$s,
with open boundaries, and without collisions.
Fig. \ref{f:CIesc:FD} presents the diffusion and convection coefficients at
$ t = 0.7\, $s, as functions of the energy. There now is a clear functional form of the data, both transport coefficients are of power-law shape, 
and a power law fit in the higher energy part above $ 5\, $keV, 
yields the power-law indexes $a_D = 1.62$ and $a_F = 0.90$. For energies below $5\ keV$, the two transport coefficients turn over and vanish.
We again insert them in the form of the fit into the FP equation, eq.~\eqref{diff}, and
solve the latter numerically. 
Fig.~\ref{f:CIesc:FP} shows the resulting energy distribution, 
together with the energy distribution from the lattice gas model. There again is a large discrepancy between the FP solution and the particles' energy distribution, the FP solution here is of extended power-law shape, yet with a slope clearly different from the one of the particles' distribution.

In contrast to the electric field acceleration, the estimate of the transport coefficients and the fit to them in Fig.~\ref{f:CIesc:FD} 
do not show obvious signs that the FP approach might fail. The distribution of energy increments is shown in Fig.~\ref{f:CIesc:p_w}, it roughly is of the same large extent as in the case of electric field acceleration,  
it exhibits a power-law part between $5\,$ and $50\,$keV, with index 
$-1.1$, and there is a turnover towards the highest energies. 
The large extent of the distribution and the at least partial power-law scaling indicate that also here the acceleration process is closer in nature to a random walk with Levy flights than to classical Brownian motion.

\subsubsection{Fractional transport equation approach}

\begin{figure}[ht]
    \sidesubfloat[]{\includegraphics[width=0.41\columnwidth]{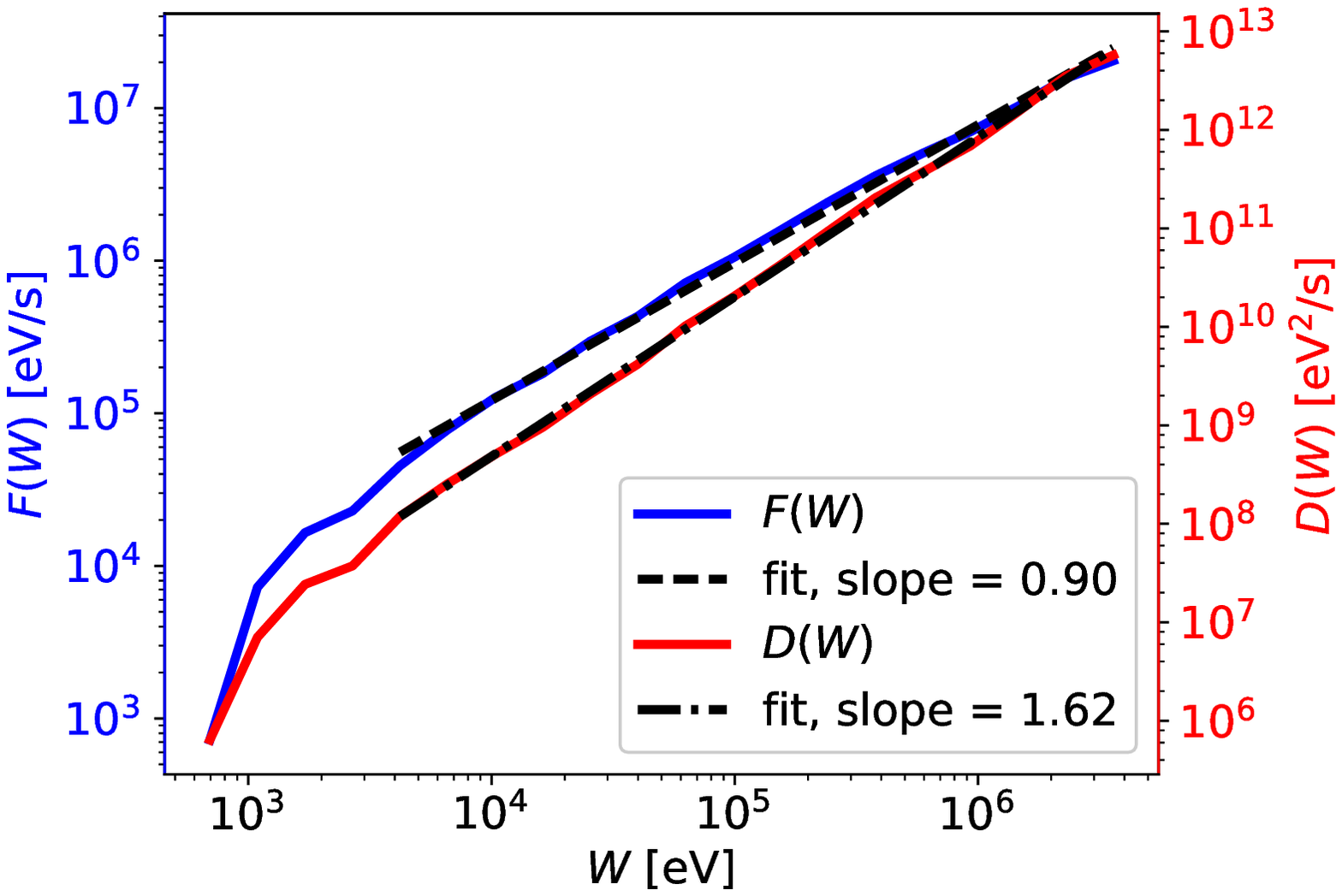}%
        \label{f:CIesc:FD}}\hfill
	\sidesubfloat[]{\includegraphics[width=0.41\columnwidth]{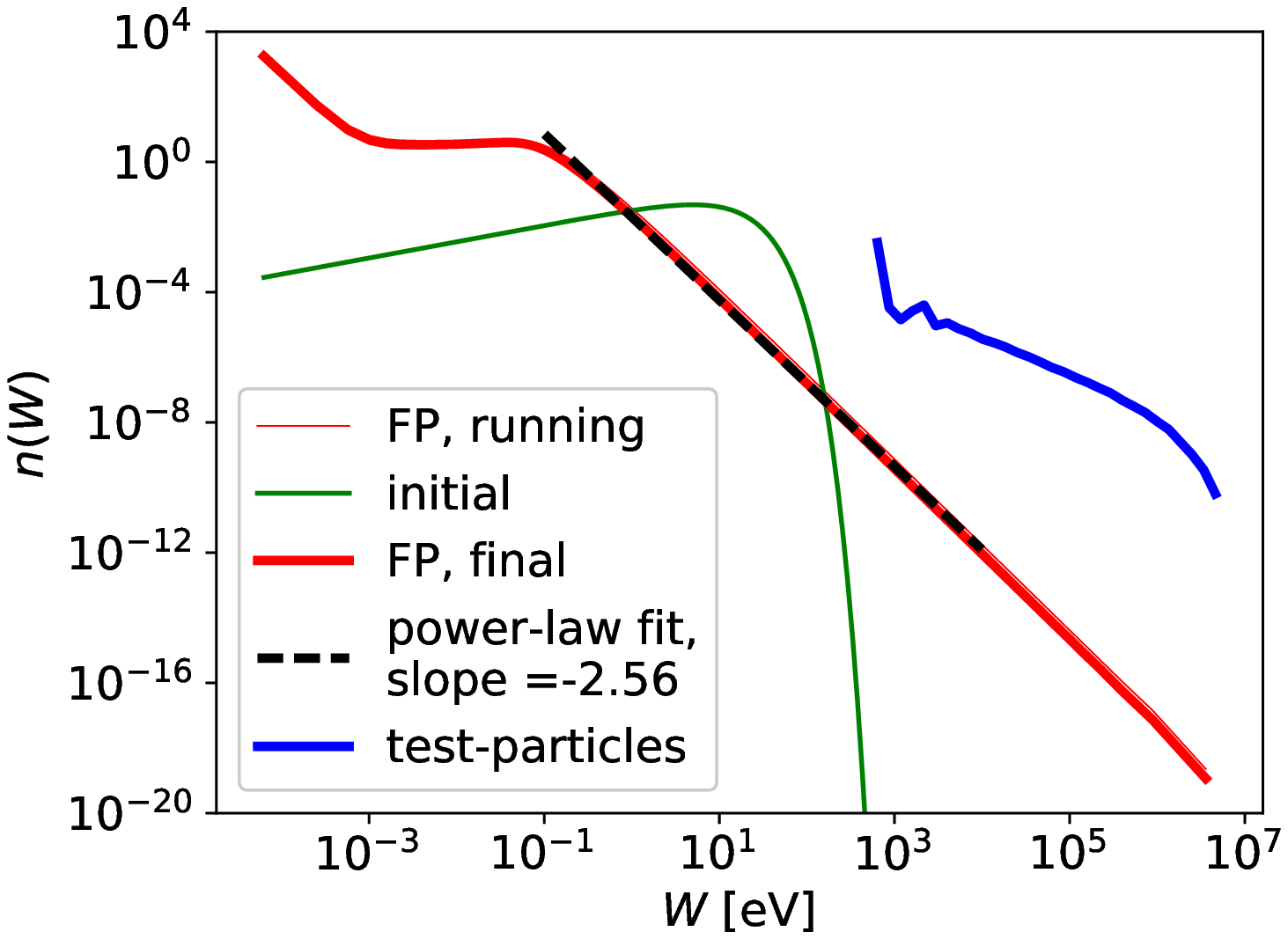}%
        \label{f:CIesc:FP}}\\
	\sidesubfloat[]{\includegraphics[width=0.41\columnwidth]{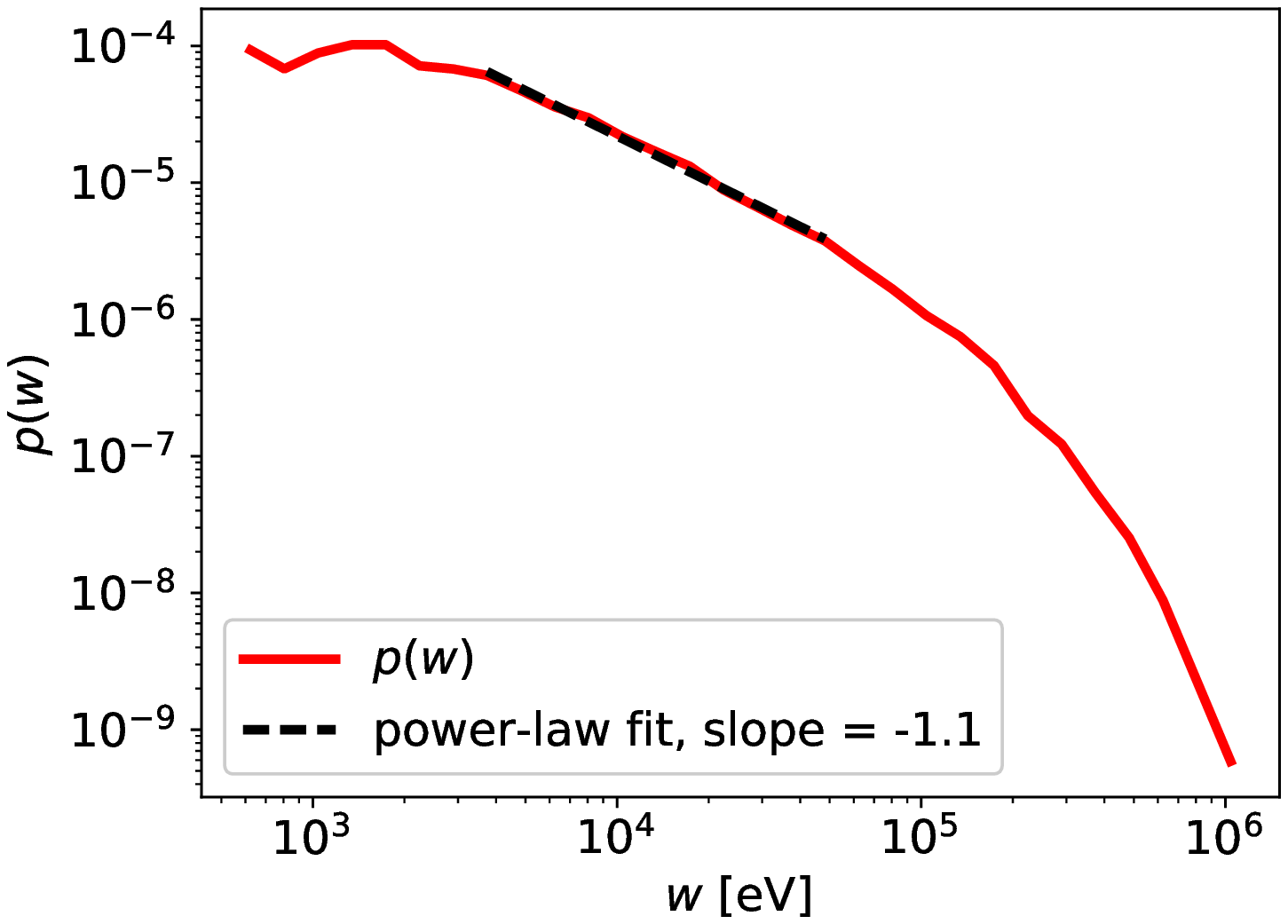}%
        \label{f:CIesc:p_w}}\hfill
	\sidesubfloat[]{\includegraphics[width=0.41\columnwidth]{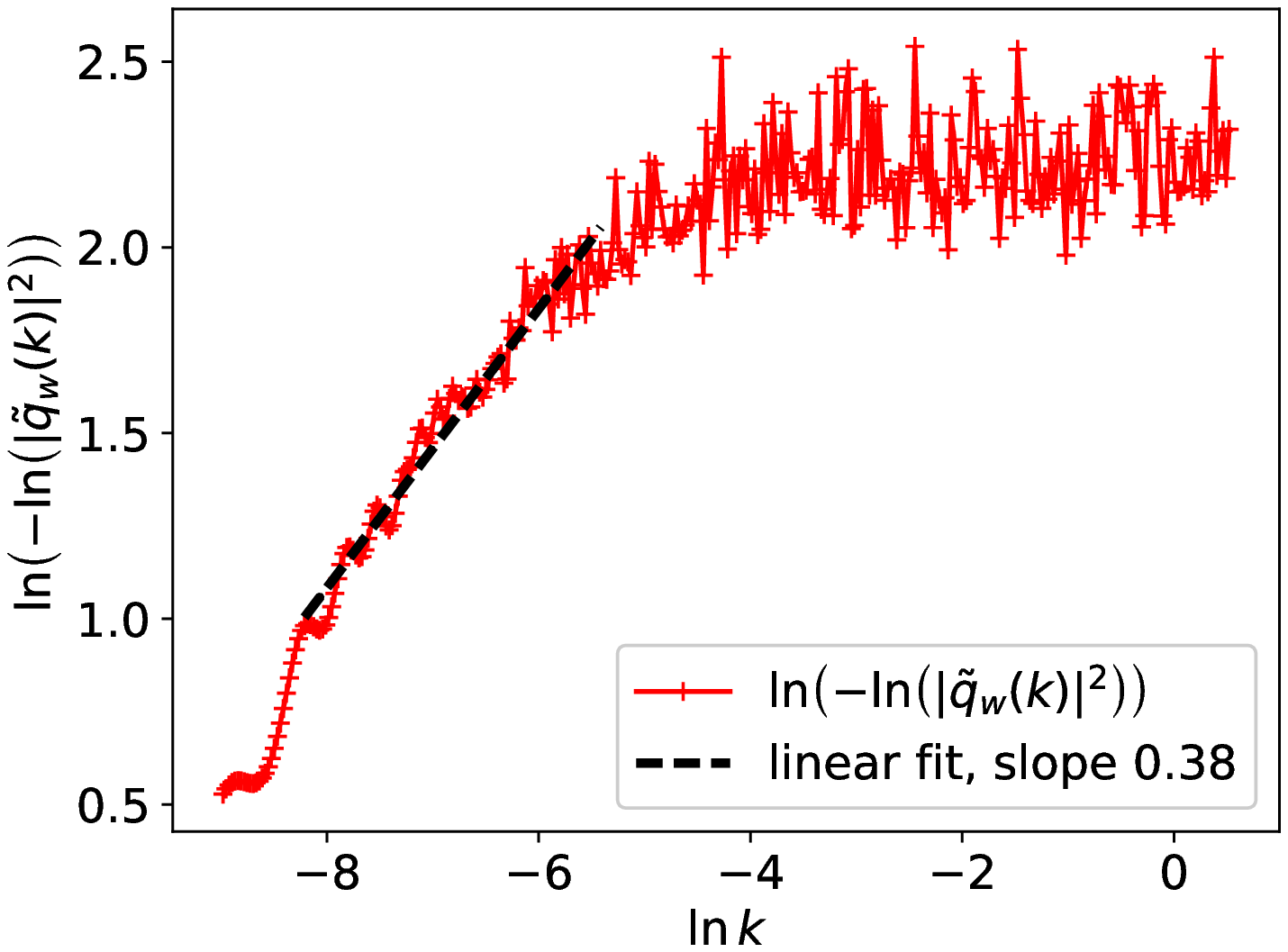}%
        \label{f:CIesc:char_fun}}\\
	\sidesubfloat[]{\includegraphics[width=0.45\columnwidth]{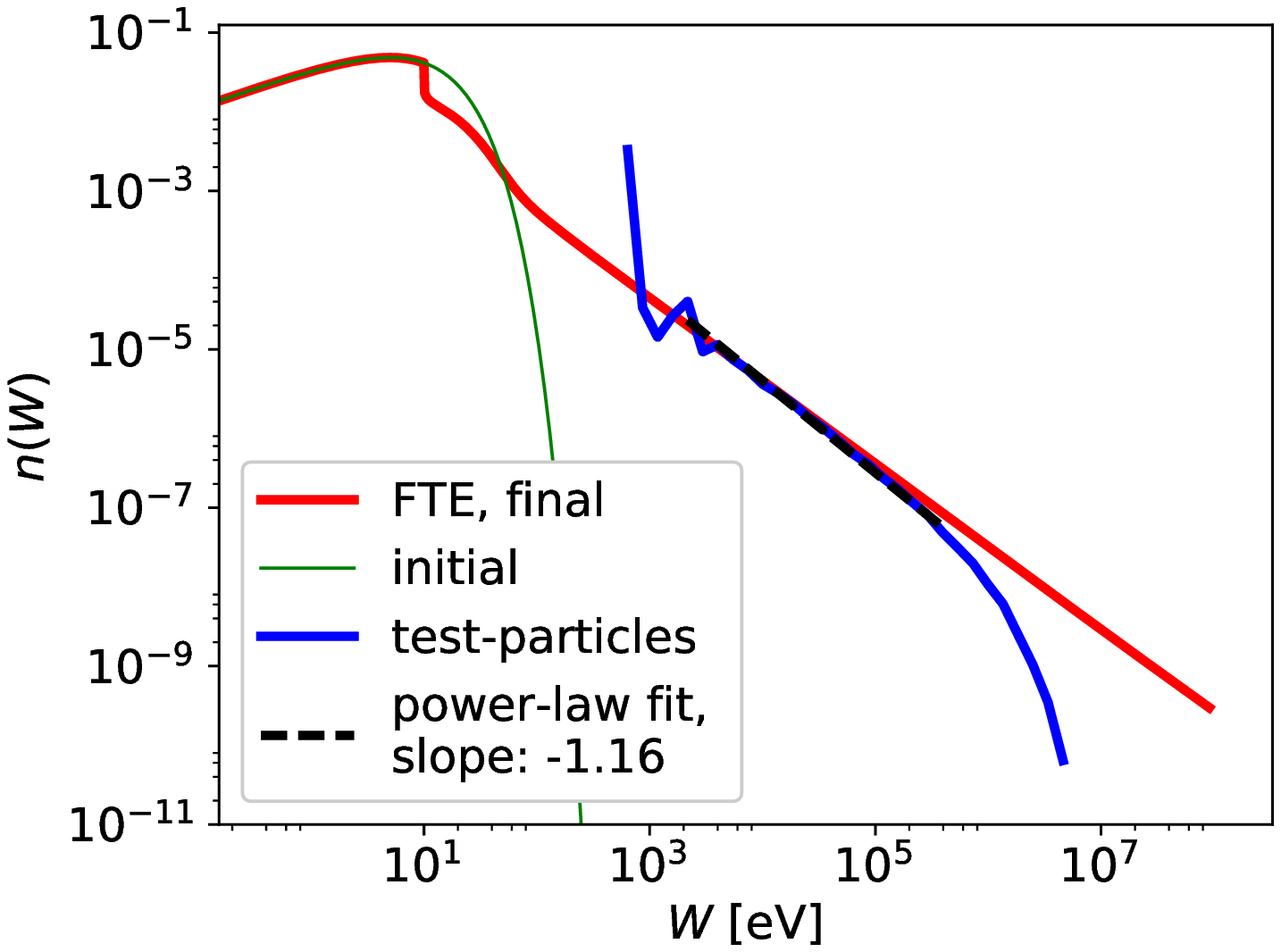}%
        \label{f:CIesc:FTE}}%
    \caption{
    \textit{Acceleration at contracting islands:}
    \protect\subref{f:CIesc:FD} Convective $F$ and diffusive $D$ transport coefficients; 
    \protect\subref{f:CIesc:FP} solution of the FP equation, together with the energy distribution from the lattice gas simulation;  \protect\subref{f:CIesc:p_w} the distribution $p(w)$ of energy increments $w$; 
    \protect\subref{f:CIesc:char_fun} the estimator of the characteristic function; 
    \protect\subref{f:CIesc:FTE} solution of the FTE, together with the energy distribution from the lattice gas simulation.}\label{f:CIesc:transport}
\end{figure}

We thus turn to the question whether the FTE is appropriate as a transport model also in the case of acceleration at contracting islands.
The order of the fractional derivative $\alpha$ can either be determined from the index $z$ of the power-law tail of $p_w(w)$
in Fig.~\ref{f:CIesc:p_w}, which yields $\alpha = -z-1 = 0.1$,
or by the 
characteristic function method [see Section \ref{ss:FTEtheor} and \eqref{e:char_fun}], which yields $\alpha = 0.38$ and $a = 29.8$ for the scale parameter. 
The characteristic function method is known in the literature to be not very precise
for low values of $\alpha$ \cite[i.e. close to 1, see e.g.][]{Borak05}, 
so that we will use here the value of $\alpha=0.1$ as inferred from the index $z$ of the power-law
tail of the increment distribution, and, as before, 
we will consider the value of $a$ just indicative.

Fig.~\ref{f:CIesc:FTE} shows the numerical solution of the FTE at time $t=0.7\,$s, together with the particles' energy distribution. The FTE also here successfully reproduces the power-law part of the particles' distribution
(for the figure we have used a value of $ a = 1.66 $).

\section{Summary and Discussion}\label{s:discussion}

It is well documented in the current literature that the well known non-linear MHD structures, i.e.\ Unstable Current Sheets, strong turbulence, and shocks, will ultimately drive a turbulent reconnection environment. In this article, we have analyzed the systematic acceleration of particles in a large scale turbulent reconnection environment, either by the electric fields associated with the reconnection sites or by reflection at contracting islands.  We have shown that the energy distribution of the particles is always a power law above a certain energy. The power law index depends on the characteristics of the interaction of the UCS with the particles, e.g.\ through the index of the power law distributed fluctuating magnetic field $\delta B$,  on the global properties of the magnetic topology, e.g.\ through the parameter $\lambda_{sc}$, and on the possible trapping of the particles, which will increase the escape time from the acceleration volume.  
For the electric field acceleration in the open simulation box the power law index is 1.7, and for the electric field acceleration with periodic boundary conditions (trapped particles), or acceleration by reflection at contracting islands, the power law index reaches asymptotically the value of 1. Varying the initial temperature of the particles in the range from $10$ to $100\,$eV does not affect the extent or the index of the power-law formed. Both acceleration processes  are extremely fast (of the order of milliseconds for the solar corona), the  released energy is mainly absorbed by the energetic tail (the heating of the bulk distribution is not important), and collisions play only a minor role at low energies. For the open simulation box, the escape time for the low energy particles increases with the escape energy, and it reaches a plateau for the high energy particles. 
The acceleration time and the escape time are related with the mean free path in-between the scatterings of the particles off the UCSs. 
\textbf{
For the particles accelerated by the electric fields in an open simulation box, the statistical properties of the UCSs \cite[see][]{Zhdankin13} play a crucial role and determine the value of the power-law index of the energy distribution.
}

Our most important finding is that the applicability of the classical FP equation breaks down. This is manifested in the inability of the FP equation to reproduce the test-particles' energy distribution, and in the practical difficulties of the expressions for $F$ and $D$ in Eqs. (4) and (3) to yield meaningful transport coefficients. The reason is that the transport in energy space is highly anomalous ('strange'), the particles move in energy space with increments that follow a power-law distribution, i.e.\ they perform Levy flights. We have shown that in such cases of anomalous transport a fractional transport equation, as introduced in \cite{Isliker17}, is appropriate, and it is indeed successful in reproducing the observed power-law distributions in energy.

The statistical analysis of the simulation data, in particular of the distribution of increments, allows deciding whether a classical FP equation or a FTE is appropriate. In the case of normal transport, the estimate of the classical transport coefficients is based on the data. In the case of anomalous transport, there are several possibilities to statistically analyze the data. In principle, a stable Levy distribution could be fitted to the distribution of increments, in order to determine the parameters of the FTE, which would have to be done in Fourier space, since the stable Levy distributions are known in analytical form only in Fourier space. The order of the fractional derivative is also directly given by the index of the power-law tail of the energy increments, and thus it is straightforward to estimate. For the other parameters, such as the scale parameter $a$, we have used the characteristic function method, which is not very accurate, and a better method is needed. After all, the form of the FTE and its parameters, most prominently the order of the fractional derivative, are directly inferred from the simulation data. We also note that the kind of data needed for this analysis can also be made available from PIC simulations and from test-particles that are tracked in the fields of MHD simulations \cite[as in][]{Isliker17}.

In the use of the FTE, we have made two simplifying assumptions, we considered only symmetric and one-sided distributions of increments, and for the time-stepping (waiting times), we considered a fixed, small time-interval, as it corresponds to the sampling of the particle evolution that we applied, and which leads to an ordinary derivative in time-direction. These assumptions did obviously not affect the success of the FTE approach we presented here.  We also note that, from its derivation, the FTE is a tool to model the high energy part (tail) of the distribution, the modeling of the low energy part could possibly be achieved by combining the fractional term in the FTE with classical diffusive and convective terms, which we will address in future work.
Moreover, spatial transport could play an important role in the acceleration process, see e.g.\ \cite{Bian17}, and also Figs.\ 
\ref{f:EFesc:tW} and \ref{f:EFesc:kicksW} give some indication 
that the spatial diffusion is of a rather complex nature.

Turbulent reconnection is a very efficient acceleration mechanism, yet 
it is not relevant for the heating of the ambient plasma. In most astrophysical sources where particle acceleration is observed, the plasma is also heated impulsively. In a recent review on turbulent reconnection, \cite{Karimabadi03b} pointed out that ``intermittent plasma turbulence will in general consist of both coherent structures and waves.'' They claim that the excitation of waves (or large scale density fluctuations) is due to the motion of the coherent structures. We propose that the combination of UCSs with large scale magnetic and density  fluctuations that are generated by the motion of the UCSs can heat the plasma by stochastic Fermi energization \cite[see][]{Pisokas16, Kontar17} and it can efficiently accelerate the particles by the systematic Fermi acceleration discussed in this article.

\begin{acknowledgements}
	This work was supported by (a) the
	national Programme for the Controlled Thermonuclear
	Fusion, Hellenic Republic, (b) the EJP Cofund Action
	SEP-210130335 EUROfusion.
	The sponsors do not bear any responsibility for the
	content of this work.
\end{acknowledgements}
\bibliographystyle{aasjournal}
\bibliography{vlahosastro}

\begin{thebibliography}{}
\expandafter\ifx\csname natexlab\endcsname\relax\def\natexlab#1{#1}\fi

\bibitem[{{Achterberg}(1981)}]{Achterberg81}
{Achterberg}, A. 1981, \aap, 97, 259

\bibitem[{{Achterberg}(1990)}]{Achterberg90b}
---. 1990, \aap, 231, 251

\bibitem[{Ambrosiano {et~al.}(1988)Ambrosiano, Matthaeus, Goldstein, \&
  Plante}]{Ambrosiano88}
Ambrosiano, J., Matthaeus, W.~H., Goldstein, M.~L., \& Plante, D. 1988, Journal
  of Geophysical Research: Space Physics, 93, 14383

\bibitem[{{Anastasiadis} \& {Vlahos}(1991)}]{Anastasiadis91}
{Anastasiadis}, A., \& {Vlahos}, L. 1991, \aap, 245, 271

\bibitem[{{Anastasiadis} \& {Vlahos}(1994)}]{Anastasiadis94}
---. 1994, \apj, 428, 819

\bibitem[{{Arzner} {et~al.}(2006){Arzner}, {Knaepen}, {Carati}, {Denewet}, \&
  {Vlahos}}]{Arzner06}
{Arzner}, K., {Knaepen}, B., {Carati}, D., {Denewet}, N., \& {Vlahos}, L. 2006,
  \apj, 637, 322

\bibitem[{Arzner \& Vlahos(2004)}]{Arzner04}
Arzner, K., \& Vlahos, L. 2004, The Astrophysical Journal Letters, 605, L69

\bibitem[{{Bian} \& {Browning}(2008)}]{Bian08}
{Bian}, N.~H., \& {Browning}, P.~K. 2008, \apjl, 687, L111

\bibitem[{Bian {et~al.}(2017)Bian, Emslie, \& Kontar}]{Bian17}
Bian, N.~H., Emslie, A.~G., \& Kontar, E.~P. 2017, The Astrophysical Journal,
  835, 262

\bibitem[{{Biskamp} \& {Welter}(1989)}]{Biskamp89}
{Biskamp}, D., \& {Welter}, H. 1989, Physics of Fluids B, 1, 1964

\bibitem[{{Borak, Szymon} {et~al.}(2005){Borak, Szymon}, {Härdle, Wolfgang},
  \& {Weron, Rafał}}]{Borak05}
{Borak, Szymon}, {Härdle, Wolfgang}, \& {Weron, Rafał}. 2005, in Statistical
  {Tools} for {Finance} and {Insurance}, cizek, pavel, härdle, wolfgang karl,
  weron, rafał (eds.) edn. (Berlin Heidelberg: Springer), 21--44

\bibitem[{Boyd(2001)}]{Boyd01}
Boyd, J.~P. 2001, Chebyshev and {Fourier} spectral methods, 2nd edn. (New York:
  Dover Publications)

\bibitem[{{Burgess} {et~al.}(2016){Burgess}, {Gingell}, \&
  {Matteini}}]{Burgers16}
{Burgess}, D., {Gingell}, P.~W., \& {Matteini}, L. 2016, \apj, 822, 38

\bibitem[{{Caprioli} \& {Spitkovsky}(2014{\natexlab{a}})}]{Caprioli14a}
{Caprioli}, D., \& {Spitkovsky}, A. 2014{\natexlab{a}}, \apj, 783, 91

\bibitem[{{Caprioli} \& {Spitkovsky}(2014{\natexlab{b}})}]{Caprioli14b}
---. 2014{\natexlab{b}}, \apj, 794, 46

\bibitem[{{Caprioli} \& {Spitkovsky}(2014{\natexlab{c}})}]{Caprioli14c}
---. 2014{\natexlab{c}}, \apj, 794, 47

\bibitem[{Cargill {et~al.}(2012)Cargill, Vlahos, Baumann, Drake, \&
  Nordlund}]{Cargill12}
Cargill, P., Vlahos, L., Baumann, G., Drake, J., \& Nordlund, {\AA}. 2012,
  Space science reviews, 173, 223

\bibitem[{{Dahlin} {et~al.}(2015){Dahlin}, {Drake}, \& {Swisdak}}]{Dahlin15}
{Dahlin}, J.~T., {Drake}, J.~F., \& {Swisdak}, M. 2015, Physics of Plasmas, 22,
  100704

\bibitem[{{de Gouveia dal Pino} \& {Lazarian}(2005)}]{delPino05}
{de Gouveia dal Pino}, E.~M., \& {Lazarian}, A. 2005, \aap, 441, 845

\bibitem[{{Decker}(1988)}]{Decker88}
{Decker}, R.~B. 1988, \ssr, 48, 195

\bibitem[{Dmitruk {et~al.}(2003)Dmitruk, Matthaeus, Seenu, \&
  Brown}]{Dmitruk03}
Dmitruk, P., Matthaeus, W., Seenu, N., \& Brown, M.~R. 2003, The Astrophysical
  Journal Letters, 597, L81

\bibitem[{{Dmitruk} {et~al.}(2004){Dmitruk}, {Matthaeus}, \&
  {Seenu}}]{Dmitruk04}
{Dmitruk}, P., {Matthaeus}, W.~H., \& {Seenu}, N. 2004, \apj, 617, 667

\bibitem[{{Drake} {et~al.}(2010){Drake}, {Opher}, {Swisdak}, \&
  {Chamoun}}]{Drake10}
{Drake}, J.~F., {Opher}, M., {Swisdak}, M., \& {Chamoun}, J.~N. 2010, \apj,
  709, 963

\bibitem[{{Drake} {et~al.}(2006){Drake}, {Swisdak}, {Che}, \& {Shay}}]{Drake06}
{Drake}, J.~F., {Swisdak}, M., {Che}, H., \& {Shay}, M.~A. 2006, \nat, 443, 553

\bibitem[{{Drake} {et~al.}(2013){Drake}, {Swisdak}, \& {Fermo}}]{Drake13}
{Drake}, J.~F., {Swisdak}, M., \& {Fermo}, R. 2013, \apjl, 763, L5

\bibitem[{{Drury}(1983)}]{Drury83}
{Drury}, L.~O. 1983, Reports on Progress in Physics, 46, 973

\bibitem[{{Fermi}(1949)}]{Fermi49}
{Fermi}, E. 1949, Physical Review, 75, 1169

\bibitem[{{Fermi}(1954)}]{Fermi54}
---. 1954, \apj, 119, 1

\bibitem[{{Galsgaard} \& {Nordlund}(1996)}]{Galsgaard96}
{Galsgaard}, K., \& {Nordlund}, {\AA}. 1996, \jgr, 101, 13445

\bibitem[{{Gardiner}(2009)}]{Gardiner09}
{Gardiner}, C. 2009, Stochastic {{Methods}}: {{A Handbook}} for the {{Natural}}
  and {{Social Sciences}} (Berlin \& Heidelberg: {Springer})

\bibitem[{Gillespie(1996)}]{Gillespie1996}
Gillespie, D.~T. 1996, Physical review E, 54, 2084

\bibitem[{{Gordovskyy} \& {Browning}(2011)}]{Gordovskyy11}
{Gordovskyy}, M., \& {Browning}, P.~K. 2011, \apj, 729, 101

\bibitem[{Guo {et~al.}(2014)Guo, Li, Daughton, \& Liu}]{Guo2014}
Guo, F., Li, H., Daughton, W., \& Liu, Y.-H. 2014, Physical Review Letters,
  113, doi:10.1103/PhysRevLett.113.155005

\bibitem[{{Guo} {et~al.}(2015){Guo}, {Liu}, {Daughton}, \& {Li}}]{Guo15}
{Guo}, F., {Liu}, Y.-H., {Daughton}, W., \& {Li}, H. 2015, \apj, 806, 167

\bibitem[{{Hoshino}(2012)}]{Hoshino12b}
{Hoshino}, M. 2012, Physical Review Letters, 108, 135003

\bibitem[{{Hoshino} \& {Lyubarsky}(2012)}]{Hoshino12}
{Hoshino}, M., \& {Lyubarsky}, Y. 2012, \ssr, 173, 521

\bibitem[{{Hughes}(1995)}]{Hughes95}
{Hughes}, B.~D. 1995, Random {{Walks}} and {{Random Environments}}, Vol. 1:
  Random Walks (Oxford: {Clarendon Press})

\bibitem[{Isliker {et~al.}(2017)Isliker, Vlahos, \& Constantinescu}]{Isliker17}
Isliker, H., Vlahos, L., \& Constantinescu, D. 2017, Physical Review Letters,
  119, 045101

\bibitem[{{Karimabadi} \& {Lazarian}(2013)}]{Karibabadi2013c}
{Karimabadi}, H., \& {Lazarian}, A. 2013, Physics of Plasmas, 20, 112102

\bibitem[{Karimabadi {et~al.}(2013)Karimabadi, Roytershteyn, Daughton, \&
  Liu}]{Karimabadi13}
Karimabadi, H., Roytershteyn, V., Daughton, W., \& Liu, Y.-H. 2013, Space
  Science Reviews, 178, 307

\bibitem[{{Karimabadi} {et~al.}(2013){Karimabadi}, {Roytershteyn}, {Wan},
  {Matthaeus}, {Daughton}, {Wu}, {Shay}, {Loring}, {Borovsky}, {Leonardis},
  {Chapman}, \& {Nakamura}}]{Karimabadi03b}
{Karimabadi}, H., {Roytershteyn}, V., {Wan}, M., {et~al.} 2013, Physics of
  Plasmas, 20, 012303

\bibitem[{{Karimabadi} {et~al.}(2014){Karimabadi}, {Roytershteyn}, {Vu},
  {Omelchenko}, {Scudder}, {Daughton}, {Dimmock}, {Nykyri}, {Wan}, {Sibeck},
  {Tatineni}, {Majumdar}, {Loring}, \& {Geveci}}]{Karimabadi2014}
{Karimabadi}, H., {Roytershteyn}, V., {Vu}, H.~X., {et~al.} 2014, Physics of
  Plasmas, 21, 062308

\bibitem[{{Kilbas, A.A.} {et~al.}(2006){Kilbas, A.A.}, {Srivastava, H. M.}, \&
  {Trujillo, J.J.}}]{Kilbas06}
{Kilbas, A.A.}, {Srivastava, H. M.}, \& {Trujillo, J.J.} 2006, Theory and
  {Applications} of {Fractional} {Differential} {Equations} (Amsterdam:
  Elsevier)

\bibitem[{{Kirk} \& {Dendy}(2001)}]{Kirk01}
{Kirk}, J.~G., \& {Dendy}, R.~O. 2001, Journal of Physics G Nuclear Physics,
  27, 1589

\bibitem[{Klafter {et~al.}(1987)Klafter, Blumen, \& Shlesinger}]{Klafter87}
Klafter, J., Blumen, A., \& Shlesinger, M.~F. 1987, Physical Review A, 35, 3081

\bibitem[{Klages {et~al.}(2008)Klages, Radons, \& Sokolov}]{Klages08}
Klages, R., Radons, G., \& Sokolov, I.~M. 2008, Anomalous {Transport}:
  {Foundations} and {Applications} (Weinheim, Germany: Wiley-VCH)

\bibitem[{Kontar {et~al.}(2017)Kontar, Perez, Harra, Kuznetsov, Emslie,
  Jeffrey, Bian, \& Dennis}]{Kontar17}
Kontar, E., Perez, J., Harra, L., {et~al.} 2017, Physical Review Letters, 118,
  155101

\bibitem[{{Koutrouvelis}(1980)}]{Koutrouvelis80}
{Koutrouvelis}, I.~A. 1980, Journal of the American Statistical Association,
  75, 918

\bibitem[{{Kowal} {et~al.}(2011){Kowal}, {de Gouveia Dal Pino}, \&
  {Lazarian}}]{Kowal11}
{Kowal}, G., {de Gouveia Dal Pino}, E.~M., \& {Lazarian}, A. 2011, \apj, 735,
  102

\bibitem[{{Krymskii}(1977)}]{Krymskii77}
{Krymskii}, G.~F. 1977, Akademiia Nauk SSSR Doklady, 234, 1306

\bibitem[{Kulsrud \& Ferrari(1971)}]{Kulsrud71}
Kulsrud, R.~M., \& Ferrari, A. 1971, Astrophysics and Space Science, 12, 302

\bibitem[{{Lazarian} \& {Opher}(2009)}]{Lazarian09}
{Lazarian}, A., \& {Opher}, M. 2009, \apj, 703, 8

\bibitem[{Lazarian {et~al.}(2012)Lazarian, Vlahos, Kowal, Yan, Beresnyak, \&
  Dal~Pino}]{Lazarian12}
Lazarian, A., Vlahos, L., Kowal, G., {et~al.} 2012, Space science reviews, 173,
  557

\bibitem[{{Lee} {et~al.}(2012){Lee}, {Mewaldt}, \& {Giacalone}}]{Lee2012}
{Lee}, M.~A., {Mewaldt}, R.~A., \& {Giacalone}, J. 2012, \ssr, 173, 247

\bibitem[{{Lenard} \& {Bernstein}(1958)}]{Lenard58}
{Lenard}, A., \& {Bernstein}, I.~B. 1958, Physical Review, 112, 1456

\bibitem[{{Longair}(2011)}]{Longair11}
{Longair}, M.~S. 2011, {High Energy Astrophysics} (Cambridge University Press)

\bibitem[{{Matsumoto} {et~al.}(2015){Matsumoto}, {Amano}, {Kato}, \&
  {Hoshino}}]{Matsumoto15}
{Matsumoto}, Y., {Amano}, T., {Kato}, T.~N., \& {Hoshino}, M. 2015, Science,
  347, 974

\bibitem[{{Matthaeus} {et~al.}(1984){Matthaeus}, {Ambrosiano}, \&
  {Goldstein}}]{Matthaeus84}
{Matthaeus}, W.~H., {Ambrosiano}, J.~J., \& {Goldstein}, M.~L. 1984, Physical
  Review Letters, 53, 1449

\bibitem[{{Matthaeus} \& {Lamkin}(1986)}]{Matthaeus86}
{Matthaeus}, W.~H., \& {Lamkin}, S.~L. 1986, Physics of Fluids, 29, 2513

\bibitem[{{Matthaeus} \& {Montgomery}(1980)}]{Matthaeus80}
{Matthaeus}, W.~H., \& {Montgomery}, D. 1980, Annals of the New York Academy of
  Sciences, 357, 203

\bibitem[{{Matthaeus} \& {Velli}(2011)}]{Matthaeus11}
{Matthaeus}, W.~H., \& {Velli}, M. 2011, \ssr, 160, 145

\bibitem[{{Melrose}(1994)}]{Melrose94}
{Melrose}, D.~B. 1994, \apjs, 90, 623

\bibitem[{{Miller} {et~al.}(1990){Miller}, {Guessoum}, \& {Ramaty}}]{Miller90}
{Miller}, J.~A., {Guessoum}, N., \& {Ramaty}, R. 1990, \apj, 361, 701

\bibitem[{{Miller} {et~al.}(1997){Miller}, {Cargill}, {Emslie}, {Holman},
  {Dennis}, {LaRosa}, {Winglee}, {Benka}, \& {Tsuneta}}]{Miller97}
{Miller}, J.~A., {Cargill}, P.~J., {Emslie}, A.~G., {et~al.} 1997, \jgr, 102,
  14631

\bibitem[{{Montroll} \& {Weiss}(1965)}]{Montroll65}
{Montroll}, E.~W., \& {Weiss}, G.~H. 1965, Journal of Mathematical Physics, 6,
  167

\bibitem[{{Onofri} {et~al.}(2006){Onofri}, {Isliker}, \& {Vlahos}}]{Onofri06}
{Onofri}, M., {Isliker}, H., \& {Vlahos}, L. 2006, Physical Review Letters, 96,
  151102

\bibitem[{Petrosian(2012)}]{Petrosian12}
Petrosian, V. 2012, Space science reviews, 173, 535

\bibitem[{{Pisokas} {et~al.}(2017){Pisokas}, {Vlahos}, {Isliker}, {Tsiolis}, \&
  {Anastasiadis}}]{Pisokas16}
{Pisokas}, T., {Vlahos}, L., {Isliker}, H., {Tsiolis}, V., \& {Anastasiadis},
  A. 2017, \apj, 835, 214

\bibitem[{Podlubny {et~al.}(2009)Podlubny, Chechkin, Skovranek, Chen, \&
  Vinagre~Jara}]{Podlubny09}
Podlubny, I., Chechkin, A., Skovranek, T., Chen, Y., \& Vinagre~Jara, B.~M.
  2009, Journal of Computational Physics, 228, 3137

\bibitem[{Podlubny {et~al.}(2013)Podlubny, Skovranek, Vinagre~Jara, Petras,
  Verbitsky, \& Chen}]{Podlubny13}
Podlubny, I., Skovranek, T., Vinagre~Jara, B.~M., {et~al.} 2013, Philosophical
  Transactions of the Royal Society A: Mathematical, Physical and Engineering
  Sciences, 371, 20120153

\bibitem[{{Priest}(2014)}]{Priest14}
{Priest}, E. 2014, {Magnetohydrodynamics of the Sun} ({Cambridge University
  Press})

\bibitem[{Ragwitz \& Kantz(2001)}]{Ragwitz2001}
Ragwitz, M., \& Kantz, H. 2001, Physical Review Letters, 87, 254501

\bibitem[{{Schneider}(1993)}]{Schneider93}
{Schneider}, P. 1993, \aap, 278, 315

\bibitem[{{Schure} {et~al.}(2012){Schure}, {Bell}, {O'C Drury}, \&
  {Bykov}}]{Schurel12}
{Schure}, K.~M., {Bell}, A.~R., {O'C Drury}, L., \& {Bykov}, A.~M. 2012, \ssr,
  173, 491

\bibitem[{{Servidio} {et~al.}(2011){Servidio}, {Dmitruk}, {Greco}, {Wan},
  {Donato}, {Cassak}, {Shay}, {Carbone}, \& {Matthaeus}}]{Servidio11}
{Servidio}, S., {Dmitruk}, P., {Greco}, A., {et~al.} 2011, Nonlinear Processes
  in Geophysics, 18, 675

\bibitem[{Turkmani {et~al.}(2005)Turkmani, Vlahos, Galsgaard, Cargill, \&
  Isliker}]{Turkmani05}
Turkmani, R., Vlahos, L., Galsgaard, K., Cargill, P., \& Isliker, H. 2005, The
  Astrophysical Journal Letters, 620, L59

\bibitem[{{Uritsky} {et~al.}(2010){Uritsky}, {Pouquet}, {Rosenberg}, {Mininni},
  \& {Donovan}}]{Uritsky10}
{Uritsky}, V.~M., {Pouquet}, A., {Rosenberg}, D., {Mininni}, P.~D., \&
  {Donovan}, E.~F. 2010, \pre, 82, 056326

\bibitem[{{Vlahos} {et~al.}(2004){Vlahos}, {Isliker}, \& {Lepreti}}]{Vlahos04}
{Vlahos}, L., {Isliker}, H., \& {Lepreti}, F. 2004, \apj, 608, 540

\bibitem[{{Vlahos} {et~al.}(2016){Vlahos}, {Pisokas}, {Isliker}, {Tsiolis}, \&
  {Anastasiadis}}]{Vlahos16}
{Vlahos}, L., {Pisokas}, T., {Isliker}, H., {Tsiolis}, V., \& {Anastasiadis},
  A. 2016, \apjl, 827, L3

\bibitem[{{Zank} {et~al.}(2015){Zank}, {Hunana}, {Mostafavi}, {Le Roux}, {Li},
  {Webb}, {Khabarova}, {Cummings}, {Stone}, \& {Decker}}]{Zank15}
{Zank}, G.~P., {Hunana}, P., {Mostafavi}, P., {et~al.} 2015, \apj, 814, 137

\bibitem[{{Zhdankin} {et~al.}(2013){Zhdankin}, {Uzdensky}, {Perez}, \&
  {Boldyrev}}]{Zhdankin13}
{Zhdankin}, V., {Uzdensky}, D.~A., {Perez}, J.~C., \& {Boldyrev}, S. 2013,
  \apj, 771, 124

\end{thebibliography}

\end{document}